\pdfoutput=1
\documentclass[12pt,a4paper]{article}
\usepackage{ifthen} % for conditional statements
\newboolean{pdflatex}
\setboolean{pdflatex}{true} % False for eps figures 

\newboolean{articletitles}
\setboolean{articletitles}{true} % False removes titles in references

\newboolean{uprightparticles}
\setboolean{uprightparticles}{false} %True for upright particle symbols

\usepackage{lscape}
\usepackage{appendix}
\usepackage{multirow}

\usepackage{microtype}
\usepackage{lineno}  % for line numbering during review
\usepackage{xspace} % To avoid problems with missing or double spaces after
\usepackage{graphicx}  % to include figures (can also use other packages)
\usepackage{color}
\usepackage{colortbl}
\graphicspath{{./figs/}} % Make Latex search fig subdir for figures

\usepackage{amsmath} % Adds a large collection of math symbols
\usepackage{amssymb}
\usepackage{amsfonts}
\usepackage{upgreek} % Adds in support for greek letters in roman typeset

\newcommand*\patchAmsMathEnvironmentForLineno[1]{%
\expandafter\let\csname old#1\expandafter\endcsname\csname #1\endcsname
\expandafter\let\csname oldend#1\expandafter\endcsname\csname
end#1\endcsname
 \renewenvironment{#1}%
   {\linenomath\csname old#1\endcsname}%
   {\csname oldend#1\endcsname\endlinenomath}%
}
\newcommand*\patchBothAmsMathEnvironmentsForLineno[1]{%
  \patchAmsMathEnvironmentForLineno{#1}%
  \patchAmsMathEnvironmentForLineno{#1*}%
}
\AtBeginDocument{%
\patchBothAmsMathEnvironmentsForLineno{equation}%
\patchBothAmsMathEnvironmentsForLineno{align}%
\patchBothAmsMathEnvironmentsForLineno{flalign}%
\patchBothAmsMathEnvironmentsForLineno{alignat}%
\patchBothAmsMathEnvironmentsForLineno{gather}%
\patchBothAmsMathEnvironmentsForLineno{multline}%
}

\usepackage{hyperref}    % Hyperlinks in references
\usepackage[all]{hypcap} % Internal hyperlinks to floats.

\def\lhcb {\mbox{LHCb}\xspace}
\def\ux85 {\mbox{UX85}\xspace}

\ifthenelse{\boolean{uprightparticles}}%
{

 \def\PDelta      {\ensuremath{\Delta}\xspace}                 
 \def\PXi      {\ensuremath{\Xi}\xspace}                 
 \def\PLambda      {\ensuremath{\Lambda}\xspace}                 
 \def\PSigma      {\ensuremath{\Sigma}\xspace}                 
 \def\POmega      {\ensuremath{\Omega}\xspace}                 
 \def\PUpsilon      {\ensuremath{\Upsilon}\xspace}                 
 
 \def\PB      {\ensuremath{\mathrm{B}}\xspace}                 
                  
 \def\PD      {\ensuremath{\mathrm{D}}\xspace}

 \def\PK      {\ensuremath{\mathrm{K}}\xspace}

 \def\Pi      {\ensuremath{\mathrm{i}}\xspace}

 \def\Ps      {\ensuremath{\mathrm{s}}\xspace}

}
{

 \mathchardef\PDelta="7101
 \mathchardef\PXi="7104
 \mathchardef\PLambda="7103
 \mathchardef\PSigma="7106
 \mathchardef\POmega="710A
 \mathchardef\PUpsilon="7107
                  
 \def\PB      {\ensuremath{B}\xspace}                 
                  
 \def\PD      {\ensuremath{D}\xspace}

 \def\PK      {\ensuremath{K}\xspace}

 \def\Pi      {\ensuremath{i}\xspace}

 \def\Ps      {\ensuremath{s}\xspace}

}

\def\squark    {\ensuremath{\Ps}\xspace}

\def\kaon  {\ensuremath{\PK}\xspace}
  \def\Kbar  {\kern 0.2em\overline{\kern -0.2em \PK}{}\xspace}

\def\Kz    {\ensuremath{\kaon^0}\xspace}
\def\Kzb   {\ensuremath{\Kbar^0}\xspace}
\def\KzKzb {\ensuremath{\Kz \kern -0.16em \Kzb}\xspace}
\def\Kp    {\ensuremath{\kaon^+}\xspace}
\def\Km    {\ensuremath{\kaon^-}\xspace}

\def\KpKm  {\ensuremath{\Kp \kern -0.16em \Km}\xspace}

  \def\Dbar    {\kern 0.2em\overline{\kern -0.2em \PD}{}\xspace}
\def\D       {\ensuremath{\PD}\xspace}

\def\Dz      {\ensuremath{\D^0}\xspace}
\def\Dzb     {\ensuremath{\Dbar^0}\xspace}
\def\DzDzb   {\ensuremath{\Dz {\kern -0.16em \Dzb}}\xspace}
\def\Dp      {\ensuremath{\D^+}\xspace}
\def\Dm      {\ensuremath{\D^-}\xspace}

\def\DpDm    {\ensuremath{\Dp {\kern -0.16em \Dm}}\xspace}

\def\B       {\ensuremath{\PB}\xspace}
  \def\Bbar    {\kern 0.18em\overline{\kern -0.18em \PB}{}\xspace}
\def\Bb      {\ensuremath{\Bbar}\xspace}
 
\def\Bz      {\ensuremath{\B^0}\xspace}
\def\Bzb     {\ensuremath{\Bbar^0}\xspace}

\def\Bs      {\ensuremath{\B^0_\squark}\xspace}

  \def\Y#1S{\ensuremath{\PUpsilon{(#1S)}}\xspace}% no space before {...}!

\def\Lbar {\ensuremath{\kern 0.1em\overline{\kern -0.1em\PLambda}}\xspace}

\newcommand{\decay}[2]{\ensuremath{#1\!\to #2}\xspace}         % {\Pa}{\Pb \Pc}

\def\to                 {\ensuremath{\rightarrow}\xspace}

\def\CP                {\ensuremath{C\!P}\xspace}
\def\CPT               {\ensuremath{C\!PT}\xspace}

\def\BsToKK       {\decay{\Bs}{\Kp\Km}}

\def\AT#1     {\ensuremath{A_{\mathrm{T}}^{#1}}\xspace}           % 2

\def\C#1      {\ensuremath{\mathcal{C}_{#1}}\xspace}                       % 9
\def\Cp#1     {\ensuremath{\mathcal{C}_{#1}^{'}}\xspace}                    % 7
\def\Ceff#1   {\ensuremath{\mathcal{C}_{#1}^{\mathrm{(eff)}}}\xspace}        % 9  
\def\Cpeff#1  {\ensuremath{\mathcal{C}_{#1}^{'\mathrm{(eff)}}}\xspace}       % 7
\def\Ope#1    {\ensuremath{\mathcal{O}_{#1}}\xspace}                       % 2
\def\Opep#1   {\ensuremath{\mathcal{O}_{#1}^{'}}\xspace}                    % 7

\newcommand{\tev}{\ensuremath{\mathrm{\,Te\kern -0.1em V}}\xspace}
\newcommand{\gev}{\ensuremath{\mathrm{\,Ge\kern -0.1em V}}\xspace}
\newcommand{\mev}{\ensuremath{\mathrm{\,Me\kern -0.1em V}}\xspace}
\newcommand{\kev}{\ensuremath{\mathrm{\,ke\kern -0.1em V}}\xspace}
\newcommand{\ev}{\ensuremath{\mathrm{\,e\kern -0.1em V}}\xspace}
\newcommand{\gevc}{\ensuremath{{\mathrm{\,Ge\kern -0.1em V\!/}c}}\xspace}
\newcommand{\mevc}{\ensuremath{{\mathrm{\,Me\kern -0.1em V\!/}c}}\xspace}
\newcommand{\gevcc}{\ensuremath{{\mathrm{\,Ge\kern -0.1em V\!/}c^2}}\xspace}
\newcommand{\gevgevcccc}{\ensuremath{{\mathrm{\,Ge\kern -0.1em V^2\!/}c^4}}\xspace}
\newcommand{\mevcc}{\ensuremath{{\mathrm{\,Me\kern -0.1em V\!/}c^2}}\xspace}

\def\invfb   {\ensuremath{\mbox{\,fb}^{-1}}\xspace}

\def\ps   {\ensuremath{{\rm \,ps}}\xspace}

\def\invps{\ensuremath{{\rm \,ps^{-1}}}\xspace}

\def\gsim{{~\raise.15em\hbox{$>$}\kern-.85em
          \lower.35em\hbox{$\sim$}~}\xspace}
\def\lsim{{~\raise.15em\hbox{$<$}\kern-.85em
          \lower.35em\hbox{$\sim$}~}\xspace}

\def\rad{\ensuremath{\rm \,rad}\xspace}

\def\tell1  {TELL1\xspace}
\def\ukl1   {UKL1\xspace}

\usepackage{cite} % Allows for ranges in citations
\usepackage{mciteplus}

\begin{document}

%%%%%%%%%%%%%%%%%%%%%%%%%
%%%%% Title     %%%%%%%%%
%%%%%%%%%%%%%%%%%%%%%%%%%
\renewcommand{\thefootnote}{\fnsymbol{footnote}}
\setcounter{footnote}{1}

% %%%%%%% CHOOSE TITLE PAGE--------
%\onecolumn
% \input{title-LHCb-ANA}
% \input{title-LHCb-CONF}
% $Id: title-LHCb-PAPER.tex 56951 2014-06-30 13:45:01Z roldeman $
% ===============================================================================
% Purpose: LHCb-PAPER journal paper title page template
% Author: 
% Created on: 2010-09-25
% ===============================================================================

%%%%%%%%%%%%%%%%%%%%%%%%%
%%%%%  TITLE PAGE  %%%%%%
%%%%%%%%%%%%%%%%%%%%%%%%%
\begin{titlepage}
\pagenumbering{roman}

% Header ---------------------------------------------------
\vspace*{-1.5cm}
\centerline{\large EUROPEAN ORGANIZATION FOR NUCLEAR RESEARCH (CERN)}
\vspace*{1.5cm}
\hspace*{-0.5cm}
\begin{tabular*}{\linewidth}{lc@{\extracolsep{\fill}}r}
\ifthenelse{\boolean{pdflatex}}% Logo format choice
{\vspace*{-2.7cm}\mbox{\!\!\!\includegraphics[width=.14\textwidth]{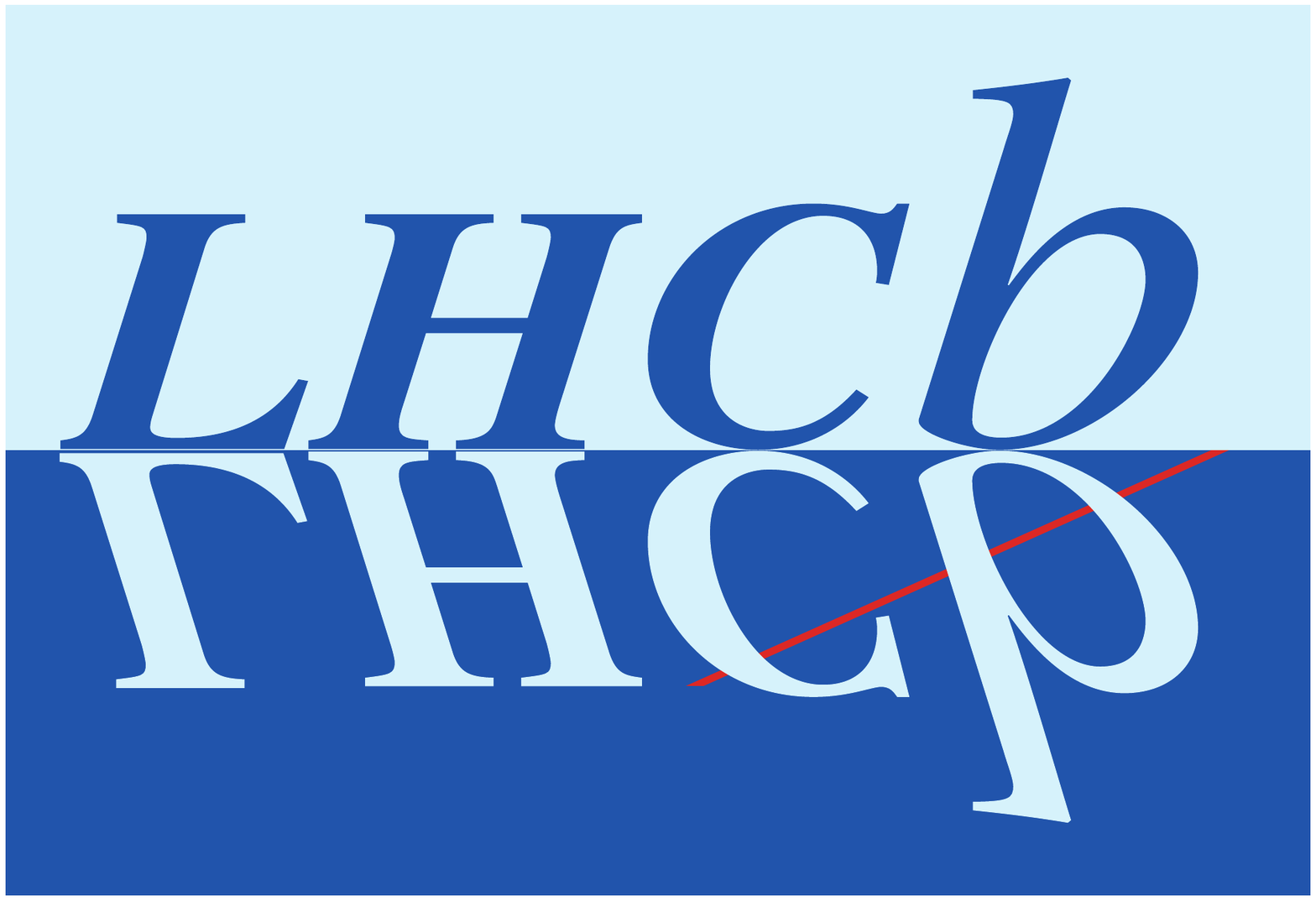}} & &}%
{\vspace*{-1.2cm}\mbox{\!\!\!\includegraphics[width=.12\textwidth]{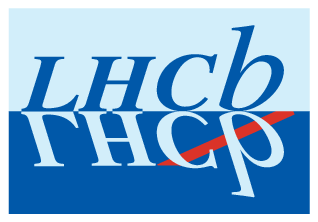}} & &}%
\\
 & & CERN-PH-EP-2014-207 \\  % ID 
 & & LHCb-PAPER-2014-045 \\  % ID 
 & & August 19, 2014 \\ % \today \\ % Date - Can also hardwire e.g.: 23 March 2010
 & & \\
% not in paper \hline
\end{tabular*}

\vspace*{2.0cm}

% Title --------------------------------------------------
{\bf\boldmath\huge
\begin{center}
Determination of $\gamma$ and $-2\beta_s$ from charmless two-body decays of\\ beauty mesons
\end{center}
}

\vspace*{2.0cm}

% Authors -------------------------------------------------
\begin{center}
%In the footnote, replace 'paper' by 'letter' in case of submission to PRL or PLB 
The LHCb collaboration\footnote{Authors are listed at the end of this paper.}
\end{center}

\vspace{\fill}

% Abstract -----------------------------------------------
\begin{abstract}
  \noindent
Using the latest LHCb measurements of time-dependent $C\!P$ violation in the $B^0_s \to K^+K^-$ decay, a U-spin relation between the decay amplitudes of $B^0_s \to K^+K^-$ and $B^0\to \pi^+\pi^-$ decay processes allows constraints to be placed on the angle $\gamma$ of the unitarity triangle and on the $B^0_s$ mixing phase $-2\beta_s$. Results from an extended approach, which uses additional inputs on $B^0\to \pi^0\pi^0$ and $B^+\to \pi^+\pi^0$ decays from other experiments and exploits isospin symmetry, are also presented. The dependence of the results on the maximum allowed amount of U-spin breaking is studied. At 68\% probability, the value $\gamma =  \left( 63.5^{\,+\, 7.2}_{\,-\,6.7} \right)^\circ~\mathrm{modulo}~180^\circ$ is determined. In an alternative analysis, the value $-2\beta_s = -0.12 ^{\,+\,0.14}_{\,-\,0.16}\,\,\mathrm{rad}$ is found. In both measurements, the uncertainties due to U-spin breaking effects up to 50\% are included.
\end{abstract}

\vspace*{2.0cm}

\begin{center}
  Submitted to Phys.~Lett.~B 
\end{center}

\vspace{\fill}

{\footnotesize 
\centerline{\copyright~CERN on behalf of the \lhcb collaboration, license \href{http://creativecommons.org/licenses/by/4.0/}{CC-BY-4.0}.}}
\vspace*{2mm}

\end{titlepage}

%%%%%%%%%%%%%%%%%%%%%%%%%%%%%%%%
%%%%%  EOD OF TITLE PAGE  %%%%%%
%%%%%%%%%%%%%%%%%%%%%%%%%%%%%%%%

%  empty page follows the title page ----
\newpage
\setcounter{page}{2}
\mbox{~}
%\newpage
%
%% Author List ----------------------------
%%  You need to get a new author list!
%\input{LHCb_authorlist.tex}
%
%The author list for journal publications is provided by the Membership Committee shortly after 'approval to go to paper' has been given.
%%It will be made available on the page
%%\verb!http://www.physik.uzh.ch/~strauman/forMemCo/LHCb-PAPER-XXXX-XXX/! .
%It will be sent to you by email shortly after a paper number has beens assigned.
%The author list should be included already at first circulation, 
%to allow new members of the collaboration to verify whether they have been included correctly.
%Occasionally a misspelled name is corrected or associated institutions become full members.
%In that case, a new author list will be sent to you.
%In case line numbering doesn't work well after including the authorlist, try moving the \verb!\bigskip! after the last author to a separate line.
%
%
%The authorship for Conference Reports should be ``The LHCb
%  collaboration'', with a footnote giving the name(s) of the contact
%  author(s), but without the full list of collaboration names.

\cleardoublepage

%\twocolumn
% %%%%%%%%%%%%% ---------

\renewcommand{\thefootnote}{\arabic{footnote}}
\setcounter{footnote}{0}

%%%%%%%%%%%%%%%%%%%%%%%%%%%%%%%%
%%%%%  Table of Content   %%%%%%
%%%%%%%%%%%%%%%%%%%%%%%%%%%%%%%%
%%%% Uncomment next 2 lines if desired
%\tableofcontents
%\cleardoublepage

%%%%%%%%%%%%%%%%%%%%%%%%%
%%%%% Main text %%%%%%%%%
%%%%%%%%%%%%%%%%%%%%%%%%%

\pagestyle{plain} % restore page numbers for the main text
\setcounter{page}{1}
\pagenumbering{arabic}

% %%%%%%% CHOOSE --------
%% ----------------------------------
%% Line numbering on the left margin 
%% ----------------------------------
%% Uncomment during review phase. 
%% Comment it out before a final submission.
%\linenumbers
%% --------------------------------
% %%%%%%%%%%%%% ---------

% You can include short sections directly in the main tex file.
% However, for larger papers it is desirable to split the text into
% several semiautonomous files, which can be revised independently.
% This is especially useful when developing a document in
% collaboration with several people, since then different parts can be
% edited independently.  This type of file organization is shown here.
% 

\section{Introduction}

The understanding of flavour dynamics is one of the most important aims of particle physics. Charge-parity ($C\!P$) violation and rare decay processes involving weak decays of $B$ mesons provide tests of the Cabibbo-Kobayashi-Maskawa (CKM) mechanism~\cite{Cabibbo:1963yz,*Kobayashi:1973fv} in the Standard Model (SM). The CKM matrix describes all flavour changing transitions of quarks in the SM. These include tree-level decays, which are expected to be largely unaffected by non-SM contributions, and flavour changing neutral current transitions characterized by the presence of loops in the relevant diagrams, which are sensitive to the presence of non-SM physics. Tests of the CKM matrix structure, commonly represented by the unitarity triangle (UT), are of fundamental importance.

Although significant hadronic uncertainties usually complicate the experimental determination of the CKM matrix elements $V_{ij}$, there are certain cases where the $V_{ij}$ can be derived with reduced or even negligible hadronic uncertainty. One of these cases involves the determination of the UT angle $\gamma$. The angle $\gamma$, defined as  $\arg \left[ - \left(V_{ud}V_{ub}^*\right) /  \left(V_{cd}V_{cb}^*\right) \right ]$, can be measured using decays that involve tree diagrams only, with almost vanishing theoretical uncertainty~\cite{Brod:2013sga}. However, $\gamma$ is experimentally the least known of the UT angles. 
World averages of the measurements performed by BaBar, Belle and LHCb~\cite{Lees:2013zd,Trabelsi:2013uj,Aaij:2013zfa,CONFGAMMA}, provided by the UTfit collaboration and CKMfitter group, are $\gamma = (70.1 \pm 7.1)^\circ$ and $\gamma = \left(68.0 ^{\,+\,8.0}_{\,-\,8.5} \right)^\circ$, respectively\footnote{The measurements of $\gamma$ are given modulo $180^\circ$ throughout this Letter.}~\cite{UTFIT,CKMFITTER}.

An alternative strategy to determine $\gamma$ using two-body charmless $B$ decays, namely $B^0 \to \pi^+\pi^-$ and $B^0_s \to K^+K^-$, has also been proposed~\cite{Fleischer:1999pa,Fleischer:2007hj,Fleischer:2010ib}. Knowledge of the $B^0$ mixing phase $2\beta$, where $\beta = \arg \left[ - \left(V_{cd}V_{cb}^*\right) /  \left(V_{td}V_{tb}^*\right) \right ]$, is needed as an input. Due to the presence of penguin diagrams in the decay amplitudes, in addition to tree diagrams, the interpretation of the observables requires knowledge of hadronic factors that cannot at present be calculated accurately from quantum chromodynamics (QCD).
However, the hadronic parameters entering the $B^0 \to \pi^+\pi^-$ and $B^0_s \to K^+K^-$ decays are related by the U-spin symmetry of strong interactions. This symmetry, related to the exchange of $d$ and $s$ quarks in the decay diagrams, can be exploited to determine the unknown hadronic factors.
A more sophisticated analysis has also been proposed~\cite{Ciuchini:2012gd}, where it is suggested to combine the U-spin analysis of $B^0 \to \pi^+\pi^-$ and $B^0_s \to K^+K^-$ decays with the isospin analysis of $B^0 \to \pi^+\pi^-$, $B^0 \to \pi^0\pi^0$ and $B^+ \to \pi^+\pi^0$ decays~\cite{Gronau:1990ka}, in order to achieve a more robust determination of $\gamma$ with respect to U-spin breaking effects. The $B^0_s$ mixing phase $-2\beta_s$, where $\beta_s = \arg \left[ - \left(V_{ts}V_{tb}^*\right) /  \left(V_{cs}V_{cb}^*\right) \right ]$, can also be determined with either analysis approach. 

An analysis based on Bayesian statistics, aimed at determining probability density functions (PDFs) for $\gamma$ and $-2\beta_s$, is presented in this Letter. This uses the latest LHCb measurements of time-dependent $C\!P$ violation in the $B^0_s \to K^+K^-$ decay, exploiting U-spin symmetry with the $B^0\to \pi^+\pi^-$ decay. An extended analysis, including measurements on $B^0\to \pi^0\pi^0$ and $B^+\to \pi^+\pi^0$ decays from other experiments, is also performed. The Letter is organized as follows. First, the theoretical formalism needed to describe \CP violation is introduced in Sec.~\ref{sec:formalism}, including the SM parameterization of the decay amplitudes of the various decays. The experimental status is given in Sec.~\ref{sec:expstatus}. In Sec.~\ref{sec:FL} we present the determination of $\gamma$ and $-2\beta_s$ using  $B^0 \to \pi^+\pi^-$ and $B^0_s \to K^+K^-$ decays, and in Sec.~\ref{sec:combined} we also add information from  $B^0\to \pi^0\pi^0$ and $B^+\to \pi^+\pi^0$ decays. The dependence of the measurements of $\gamma$ and $-2\beta_s$ on the amount of U-spin breaking is studied in detail in both cases. Finally, conclusions are drawn in Sec.~\ref{sec:conclusions}.

\section{Theoretical formalism}
\label{sec:formalism}

Assuming \CPT invariance, the \CP asymmetry as a function of decay time for a neutral $B^0$ or $B^0_s$ meson decaying to a self-conjugate final state $f$, with $f=\pi^+ \pi^-$, $\pi^0 \pi^0$ or $K^+ K^-$, is given by
\begin{equation}
\mathcal{A}(t)\equiv\frac{\Gamma_{{\Bb}^0_{(s)} \to f}(t)-\Gamma_{B^0_{(s)} \to f}(t)}{\Gamma_{{\Bb}^0_{(s)} \to f}(t)+\Gamma_{B^0_{(s)} \to f}(t)}=\frac{-C_f \cos\left(\Delta m_{d(s)} t\right) + S_f \sin\left(\Delta m_{d(s)}t\right)}{\cosh\left(\frac{\Delta\Gamma_{d(s)}}{2} t\right) + A^{\Delta\Gamma}_f \sinh\left(\frac{\Delta\Gamma_{d(s)}}{2} t\right)},
\end{equation}
where $\Delta m_{d(s)} \equiv m_{{d(s)},\,\mathrm{H}}-m_{{d(s)},\,\mathrm{L}}$ and $\Delta\Gamma_{d(s)} \equiv \Gamma_{{d(s)},\,\mathrm{L}}-\Gamma_{{d(s)},\,\mathrm{H}}$ are the mass and width differences of the $B^0_{(s)}$--$\Bb^0_{(s)}$ system mass eigenstates. The subscripts $\mathrm{H}$ and $\mathrm{L}$ denote the heavy and light eigenstates. With this convention, the value of $\Delta m_{d(s)}$ is positive by definition, and that of $\Delta\Gamma_{s}$ is measured to be positive~\cite{XIE}, $\Delta\Gamma_s = 0.106 \pm 0.011\,(\mathrm{stat})\pm0.007\,(\mathrm{syst})\invps$~\cite{Aaij:2013oba}. The value of $\Delta\Gamma_{d}$ is also positive in the SM and is expected to be much smaller than that of $\Delta\Gamma_{s}$, $\Delta\Gamma_{d} \simeq 3\times10^{-3}\invps$~\cite{UTFIT}. The quantities $C_f$, $S_f$ and $A^{\Delta\Gamma}_f$ are 
\begin{equation}
\begin{split}
C_{f} \equiv \frac{1-|\lambda_f|^2}{1+|\lambda_f|^2},\,\,\,\,\,\,\,\,\,\,\,S_{f} \equiv  \frac{2 {\rm Im} \lambda_f}{1+|\lambda_f|^2}\,\,\,\,\,\mathrm{and}\,\,\,\,\,\,A^{\Delta\Gamma}_f \equiv  - \frac{2 {\rm Re} \lambda_f}{1+|\lambda_f|^2},
\end{split}\label{eq:adirmix} 
\end{equation}
where $\lambda_f$ is given by
\begin{equation}
\lambda_f \equiv \frac{q}{p}\frac{\bar{A}_f}{A_f}.
\end{equation}
The two mass eigenstates of the effective Hamiltonian in the $B^0_{(s)}$--$\Bb^0_{(s)}$ system are $p|B^0_{(s)}\rangle \pm q|\Bb^0_{(s)}\rangle$, where $p$ and $q$ are complex parameters satisfying the relation $\left| p \right|^2 + \left| q \right|^2 = 1$. The parameter $\lambda_f$ is thus related to $B^0_{(s)}$--$\Bb^0_{(s)}$ mixing (via $q/p$) and to the decay amplitudes of the $B^0_{(s)} \to f$ decay ($A_f$) and of the $\Bb^0_{(s)} \to f$ decay ($\bar{A}_f$). Assuming negligible $C\!P$ violation in mixing ($|q/p|=1$), as expected in the SM and supported by current experimental determinations~\cite{bib:hfagbase,Aaij:2013gta}, the terms $C_{f}$ and $S_{f}$ parameterize \CP violation in the decay and in the interference between mixing and decay, respectively.
From the definitions given in Eq.~\ref{eq:adirmix}, it follows that
\begin{equation}
\left( C_{f} \right)^2 + \left( S_{f} \right)^2 + \left( A^{\Delta\Gamma}_f \right)^2 = 1.
\end{equation}
It is then possible to express the magnitude (but not the sign) of $A^{\Delta\Gamma}_f$ as a function of $C_{f}$ and $S_{f}$. There are therefore two independent parameters, which can be chosen, for example, to be ${\rm Re} \lambda_f$ and ${\rm Im} \lambda_f$, or $C_{f}$ and $S_{f}$. In the latter case, the sign of $A^{\Delta\Gamma}_f$ carries additional information.

The \CP-averaged branching fraction is given by
\begin{equation}
\mathcal{B}_{f} = \frac{1}{2} F(B^0_{(s)} \to f) \left ( \left| \bar{A}_f \right|^2 + \left| A_f \right|^2 \right),
\end{equation}
where
\begin{equation}
F(B^0 \to \pi^+\pi^-) = \frac{\sqrt{m_{B^0}^2-4m_{\pi^+}^2}}{m_{B^0}^2}\tau_{B^0},
\end{equation}
\begin{equation}
F(B^0 \to \pi^0 \pi^0) = \frac{\sqrt{m_{B^0}^2-4m_{\pi^0}^2}}{m_{B^0}^2}\tau_{B^0},
\end{equation}
\begin{equation}
F(B^0_s \to K^+K^-) = \frac{\sqrt{m_{B^0_s}^2-4m_{K^+}^2}}{m_{B^0_s}^2} \left[ 2\tau_{B^0_s} - \left(1- y_s^2 \right) \tau(B^0_s \to K^+K^-) \right],\label{eq:fleischerlifetime}
\end{equation}
with $\tau_{B^0}\equiv 1/\Gamma_d$, $\tau_{B^0_s} \equiv 1/\Gamma_s$ and $y_s \equiv \Delta\Gamma_s / (2\Gamma_s)$. The term $m_x$ is the mass of the meson $x$, $\Gamma_{d(s)} \equiv (\Gamma_{{d(s)},\,\mathrm{L}}+\Gamma_{{d(s)},\,\mathrm{H}})/2$ is the average decay width of the $B^0_{(s)}$ meson, and  $\tau(B^0_s \to K^+K^-)$ is the effective lifetime measured using $B^0_s \to K^+K^-$ decays. The extra term is Eq.~\ref{eq:fleischerlifetime} follows from the fact that the $\Bb^0_s - B^0_s$ meson system is characterized by a sizeable decay width difference. This leads to a difference between the measured (\emph{i.e.} decay-time-integrated) branching fraction and the theoretical branching fraction, and a correction is applied using the corresponding effective lifetime measurement~\cite{DeBruyn:2012wj}.

In the case of a $B^+$ meson decaying to a final state $f$, the \CP asymmetry is given by
\begin{equation}
\mathcal{A}_{f} = \frac{ \left| {\bar{A}}_{\bar{f}} \right|^2 - \left| A_f \right|^2 } { \left| {\bar{A}}_{\bar{f}} \right|^2 + \left| A_f \right|^2 },
\end{equation}
and the \CP-averaged branching fraction is
\begin{equation}
\mathcal{B}_{f} = \frac{1}{2} F(B^+ \to f) \left ( \left| {\bar{A}}_{\bar{f}} \right|^2 +\left| A_f \right|^2 \right),
\end{equation}
where
\begin{equation}
F(B^+ \to \pi^+\pi^0) = \frac{\sqrt{m_{B^+}^2-(m_{\pi^+}+m_{\pi^0})^2}}{m_{B^+}^2}\tau_{B^+},
\end{equation} 
with $\tau_{B^+}$ the lifetime and $m_{B^+}$ the mass of the $B^+$ meson.

Adopting the parameterization from Ref.~\cite{Fleischer:1999pa} and its extension from Ref.~\cite{Ciuchini:2012gd}, assuming isospin symmetry and neglecting electroweak penguin contributions, the following expressions for the various $C\!P$ asymmetry terms and branching fractions are obtained in the framework of the SM
\begin{equation}
C_{\pi^+\pi^-}=-\frac{2d\sin(\vartheta)\sin(\gamma)}{1-2d\cos(\vartheta)\cos(\gamma)+d^{2}},\label{eq:cpipi}
\end{equation}
\begin{equation}
S_{\pi^+\pi^-}=-\frac{\sin(2\beta+2\gamma)-2d\cos(\vartheta)\sin(2\beta+\gamma)+d^{2}\sin(2\beta)}
{1-2d\cos(\vartheta)\cos(\gamma)+d^{2}},\label{eq:spipi}
\end{equation}
\begin{equation}
C_{\pi^0\pi^0}=-\frac{2dq\sin(\vartheta_q-\vartheta)\sin(\gamma)}{q^2+2dq\cos(\vartheta_q-\vartheta)\cos(\gamma)+d^{2}},\label{eq:cpi0pi0}
\end{equation}
\begin{equation}
\mathcal{A}_{\pi^+\pi^0}=0,\label{eq:cpipi0}
\end{equation}
\begin{equation}
C_{K^+K^-}=\frac{2\tilde{d}^{\prime}\sin(\vartheta^{\prime})\sin(\gamma)}{1+2\tilde{d}^{\prime}\cos(\vartheta^{\prime})\cos(\gamma)+\tilde{d}^{\prime2}},\label{eq:ckk}
\end{equation}
\begin{equation}
S_{K^+K^-}=-\frac{\sin(-2\beta_s+2\gamma)+2\tilde{d}^{\prime}\cos(\vartheta^{\prime})\sin(-2\beta_s+\gamma)+\tilde{d}^{\prime2}\sin(-2\beta_s)}
{1+2\tilde{d}^{\prime}\cos(\vartheta^{\prime})\cos(\gamma)+\tilde{d}^{\prime2}},\label{eq:skk}
\end{equation}
\begin{equation}
\mathcal{B}_{\pi^+\pi^-}=F(B^0 \to \pi^+\pi^-)|D|^2(1-2d\cos(\vartheta)\cos(\gamma)+d^{2}),\label{eq:bpipi}
\end{equation}
\begin{equation}
\mathcal{B}_{\pi^0\pi^0}=F(B^0 \to \pi^0\pi^0)\frac{|D|^2}{2}(q^2+2dq\cos(\vartheta_q-\vartheta)\cos(\gamma)+d^{2}),\label{eq:bpi0pi0}
\end{equation}
\begin{equation}
\mathcal{B}_{\pi^+\pi^0}=F(B^+ \to \pi^+\pi^0)\frac{|D|^2}{2}(1+q^2+2q\cos(\vartheta_q)),\label{eq:bpipi0}
\end{equation}
\begin{equation}
\mathcal{B}_{K^+K^-}=F(B^0_s \to K^+K^-) \frac{\lambda^2}{(1-\lambda^2/2)^2}|D^{\prime}|^2(1+2\tilde{d}^{\prime}\cos(\vartheta^{\prime})\cos(\gamma)+\tilde{d}^{\prime2}),\label{eq:bkk}
\end{equation}
where $\tilde{d}^{\prime} \equiv d^\prime(1-\lambda^2)/\lambda^2$ and $\lambda \equiv | V_{us}| / \sqrt{| V_{ud}|^2 +| V_{us}|^2 }$.
In addition, $A^{\Delta\Gamma}_{K^+K^-}$ can be expressed as
\begin{equation}
A^{\Delta\Gamma}_{K^+K^-}=-\frac{\cos(-2\beta_s+2\gamma)+2\tilde{d}^{\prime}\cos(\vartheta^{\prime})\cos(-2\beta_s+\gamma)+\tilde{d}^{\prime2}\cos(-2\beta_s)}
{1+2\tilde{d}^{\prime}\cos(\vartheta^{\prime})\cos(\gamma)+\tilde{d}^{\prime2}}.\label{eq:adgammakk}
\end{equation}
The quantities $|D|$, $d$, $\vartheta$, $q$ and $\vartheta_q$ are real-valued hadronic parameters related to the decay amplitudes of $B^0 \to \pi^+\pi^-$, $B^0 \to \pi^0\pi^0$  and $B^+ \to \pi^+\pi^0$ decays, whereas $|D^{\prime}|$, $d^{\prime}$ and $\vartheta^{\prime}$ are the analogues of $|D|$, $d$ and $\vartheta$ for the $B^0_s \to K^+K^-$ decay. They are defined as
\begin{equation}
D^{(\prime)} \equiv A \lambda^3 R_u \left( -\mathrm{T}^{(\prime)} -\mathrm{P}^{(\prime) u} + \mathrm{P}^{(\prime) t} \right),
\end{equation}
\begin{equation}
d^{(^\prime)} e^{i\vartheta^{(^\prime)}} \equiv \frac{1}{R_u} \frac{\mathrm{P}^{(^\prime) c}  - \mathrm{P}^{(^\prime)t}}{\mathrm{T}^{(^\prime)} +\mathrm{P}^{(^\prime) u} - \mathrm{P}^{(^\prime) t} },
\end{equation}
\begin{equation}
q e^{i\vartheta_q} \equiv \frac{\mathrm{C} - \mathrm{P}^u  +\mathrm{P}^t }{\mathrm{T} +\mathrm{P}^u - \mathrm{P}^t },
\end{equation}
where $\mathrm{T}$ and $\mathrm{C}$ represent the contributions from $\bar{b} \to \bar{u}W^+(\to u\bar{d})$ tree and colour-suppressed tree transitions, $\mathrm{P}^q$ represents the contributions from $\bar{b} \to \bar{d} g(\to \bar{u}u)$ or $\bar{b} \to \bar{d} g(\to \bar{d}d)$ penguin transitions (the index $q 
\in \{u,\,c,\,t\}$ indicates the flavour of the internal quark in the penguin loop), $R_{u}$ is one of the sides of the UT
\begin{equation}
R_{u} = \frac{1}{\lambda}\left(1-\frac{\lambda^{2}}{2}\right)\left|\frac{V_{ub}}{V_{cb}}\right|,
\end{equation}
and $A \equiv 1/ \lambda \left | V_{cb} / V_{us} \right |$. 
Analogously, $\mathrm{T}^\prime$ represents the contribution from $\bar{b} \to \bar{u}W^+(\to u\bar{s})$ tree transitions, and $\mathrm{P}^{\prime q}$ represents the contributions from $\bar{b} \to \bar{s} g(\to \bar{u}u)$ penguin transitions.

\section{Experimental status}
\label{sec:expstatus}

$C\!P$ violation both in decay amplitudes and in their interference with the $\Bz-\Bzb$ mixing amplitude has been seen in $B^0 \to \pi^+\pi^-$  decays by the BaBar~\cite{Lees:2013bb} and Belle~\cite{Adachi:2013mae} experiments, which also provided measurements of $C\!P$ violation in the $B^+ \to \pi^+\pi^0$~\cite{Aubert:2007hh,Duh:2012ie} and $B^0 \to \pi^0\pi^0$~\cite{Lees:2013bb,Abe:2004mp} decays. LHCb has recently published measurements of $C\!P$ violation in $B^0\to\pi^+\pi^-$ and $B^0_s \to K^+K^-$ decays~\cite{Aaij:2013tna}. Measurements of branching fractions for $B^0 \to \pi^+\pi^-$, $B^+ \to \pi^+\pi^0$ and $B^0 \to \pi^0\pi^0$ decays have been made by BaBar~\cite{Aubert:2006fha,Aubert:2007hh,Lees:2013bb} and Belle~\cite{Duh:2012ie,Abe:2004mp}. CDF and LHCb have also measured the $B^0 \to \pi^+\pi^-$ branching fraction, as well as that of the $B^0_s \to K^+K^-$ decay~\cite{Aaltonen:2011qt,Aaij:2012as}, using the world average of the $B^0 \to K^+\pi^-$ branching fraction for normalization~\cite{bib:hfagbase}.
The current experimental knowledge is summarized in Table~\ref{tab:exp_summary}.

The LHCb measurement of $C_{K^+K^-}$ and $S_{K^+K^-}$ in Ref.~\cite{Aaij:2013tna} was obtained using the constraint
\begin{equation}
A^{\Delta\Gamma}_{K^+K^-}=-\sqrt{1- \left(C_{K^+K^-}\right)^2 - \left( S_{K^+K^-} \right)^2}
\end{equation}
in the maximum likelihood fit.
In the same analysis, the sign of $A^{\Delta\Gamma}_{K^+K^-}$ was verified to be negative, as expected in the SM. A measurement of $A^{\Delta\Gamma}_{K^+K^-}$ has also been made by LHCb via an effective lifetime measurement of the $B^0_s \to K^+K^-$ decay, using the same data sample as in Ref.~\cite{Aaij:2013tna}, but with different event selection. The result is $A^{\Delta\Gamma}_{K^+K^-}=-0.87 \pm 0.17\,(\mathrm{stat}) \pm 0.13\,(\mathrm{syst})$~\cite{LHCb-PAPER-2014-011}. In the analysis presented in this Letter, $A^{\Delta\Gamma}_{K^+K^-}$ is constrained to have a negative value.
\begin{table}[tb]
\begin{center}
\caption{\small Current knowledge of \CP violation parameters and \CP-averaged branching fractions of $B^0\to\pi^+\pi^-$, $B^0\to\pi^0\pi^0$, $B^+\to\pi^+\pi^0$  and $B^0_s \to K^+K^-$ decays from BaBar, Belle, CDF and LHCb. The parameter $\rho(X,\,Y)$ is the statistical correlation between $X$ and $Y$. The first uncertainties are statistical and the second systematic.}\label{tab:exp_summary}
\resizebox{1\textwidth}{!}{
\begin{tabular}{c|c|c|c|c}
Quantity & BaBar & Belle & CDF & LHCb \\
\hline
$C_{\pi^+\pi^-}$ &$-0.25\pm0.08\pm0.02$ & $-0.33\pm0.06\pm0.03$ & $-$ & $-0.38\pm0.15\pm0.02$ \\
$S_{\pi^+\pi^-}$ & $-0.68\pm0.10\pm0.03$ & $-0.64\pm0.08\pm0.03$  & $-$ &$-0.71\pm0.13\pm0.02$ \\
$\mathrm{\rho}(C_{\pi^+\pi^-},\,S_{\pi^+\pi^-})$ & $-0.06$ &$-0.10$ & $-$ &$~~0.38$ \\
$\mathcal{B}_{\pi^+\pi^-}\times 10^6$ & $ \phantom{-0}5.5 \pm ~0.4\phantom~ \pm 0.3\phantom{0} $ & $ \phantom{-}5.04 \pm 0.21 \pm 0.18$ & $5.02 \pm 0.33 \pm 0.35 $ & $\phantom{-}5.08 \pm 0.17 \pm 0.37$ \\
\hline
$C_{K^+K^-}$ & $ - $ & $ - $ & $ - $ & $\phantom{-}0.14\pm0.11\pm0.03$ \\
$S_{K^+K^-}$ & $ - $ & $ - $ & $ - $ & $\phantom{-}0.30\pm0.12\pm0.04$ \\
$\mathrm{\rho}(C_{K^+K^-},\,S_{K^+K^-})$ & $ - $ & $ - $ & $ - $ & $~~0.02$ \\
$\mathcal{B}_{K^+K^-}\times 10^6$ & $ - $ & $38^{\,+\,10}_{\,-\,9}\pm 7$ & $25.8 \pm 2.2 \pm 1.7$ & $\phantom{-}23.0 \pm \hspace{1mm} 0.7 \hspace{1mm} \pm 2.3\phantom{0}$ \\
\hline
$\mathcal{A}_{\pi^+\pi^0}$ & $ - 0.03 \pm 0.08 \pm 0.01 $ & $ - 0.025 \pm 0.043 \pm 0.007 $ & $ - $ & $ - $ \\
$\mathcal{B}_{\pi^+\pi^0}\times 10^6$ & $ \phantom{-}5.02 \pm 0.46 \pm 0.29 $ & $ \phantom{-}5.86\phantom{0} \pm 0.26\phantom{0} \pm 0.38\phantom{0} $ & $ - $ & $ - $ \\
\hline
$C_{\pi^0\pi^0}$ & $ -0.43 \pm 0.26 \pm 0.05 $ & $ -0.44 ^{\,+\,0.53}_{\,-\,0.52} \pm 0.17^{\phantom{A^A}} $ & $ - $ & $ - $ \\
$\mathcal{B}_{\pi^0\pi^0}\times 10^6$ & $ \phantom{-}1.83 \pm 0.21 \pm 0.13 $  & $ 2.3 ^{\,+\,0.4\,+\,0.2}_{\,-\,0.5\,-\,0.3} $  & $ - $ & $ - $ 
\end{tabular}}
\end{center}
\end{table}

\section{{Determination of \boldmath $\gamma$} and {\boldmath$-2\beta_s$} from {\boldmath $B^0 \to \pi^+\pi^-$} and {\boldmath $B^0_s \to K^+K^-$} decays}
\label{sec:FL}

A method to determine $\gamma$ and $-2\beta_s$ using \CP asymmetries and branching fractions of $B^0 \to \pi^+\pi^-$ and $B^0_s \to K^+K^-$ decays, exploiting the approximate U-spin symmetry of strong interactions, was proposed in Refs.~\cite{Fleischer:1999pa,Fleischer:2007hj,Fleischer:2010ib}. Typical U-spin breaking corrections are expected to be around the 30\% level~\cite{Gronau:2013mda,Nagashima:2007qn}. In the limit of strict U-spin symmetry, one has $d=d^\prime$, $\vartheta=\vartheta^\prime$ and $|D|=|D^\prime|$. As pointed out in Ref.~\cite{Fleischer:1999pa}, the equalities $d=d^\prime$ and $\vartheta=\vartheta^\prime$ do not receive U-spin breaking corrections within the factorization approximation, in contrast with the equality $|D|=|D^\prime|$,
\begin{equation}
\left | \frac{D^\prime}{D} \right |_{\mathrm{fact}} = \frac{f_K}{f\pi} \,\,\,\frac{f^+_{B^0_s K} (m^2_K)}{f^+_{B^0 \pi} (m^2_\pi)} \,\,\,\frac{m^2_{B^0_s}-m^2_K}{m^2_{B^0}-m^2_\pi}, 
\end{equation}
where $f_K$ and $f_\pi$ are the kaon and pion decay constants, and $f^+_{B^0_s K} (m^2_K)$ and $f^+_{B^0 \pi} (m^2_\pi)$ parameterize hadronic matrix elements. These quantities have been determined using QCD sum rules~\cite{Duplancic:2008tk}, yielding
\begin{equation}
\left | \frac{D^\prime}{D} \right |_{\mathrm{fact}} = 1.41 ^{+0.20}_{-0.11}.\label{eq:DpoverD}\nonumber
\end{equation}
To take into account non-factorizable U-spin breaking corrections, we parameterize the effect of the breaking as
\begin{equation}
\left | D^\prime \right | =  \left | \frac{D^\prime}{D} \right |_{\mathrm{fact}} \left| D \right | \ \left | 1+ r_D e^{i\vartheta_{r_D}} \right |,\label{eq:break1}
\end{equation}
\begin{equation}
d^\prime e^{i\vartheta^\prime} = d e^{i\vartheta} \frac{ 1+ r_G e^{i\vartheta_{r_G}} }{1+ r_D e^{i\vartheta_{r_D}}},\label{eq:break2}
\end{equation}
where $r_D$ and $r_G$ are relative magnitudes, and $\vartheta_{r_D}$ and $\vartheta_{r_G}$ are phase shifts caused by the breaking. 
In the absence of non-factorizable U-spin breaking, one has $r_D = 0$ and $r_G=0$.

We perform two distinct analyses, to determine either $\gamma$ or $-2\beta_s$. They are referred to as analyses A and B, respectively. To improve the precision on the determination of $\gamma$, in analysis A the value of $-2\beta_s$ is constrained as
\begin{equation}
-2\beta_s = - 2 \lambda^2 \bar{\eta} \left [ 1+\lambda^2 \left( 1-\bar{\rho} \right ) \right ],
\end{equation}
which is valid in the SM up to terms of order $\lambda^4$. The parameters $\bar{\rho}$ and $\bar{\eta}$ determine the apex of the UT, and are defined as $\bar{\rho}+i\bar{\eta}\equiv-(V_{ud}V_{ub}^*)/(V_{cd}V_{cb}^*)$. Since $\bar{\rho}$ and $\bar{\eta}$ can be written as functions of $\beta$ and $\gamma$ as
\begin{equation}
\bar{\rho} = \frac{\sin \beta \cos \gamma}{\sin(\beta+\gamma)},\,\,\,\,\,\,\,\,\,\,\,\,\,\,\,\,\,\,\,\,
\bar{\eta} = \frac{\sin \beta \sin \gamma}{\sin(\beta+\gamma)},
\end{equation}
we can express $-2\beta_s$ in terms of $\beta$ and $\gamma$. To determine $-2\beta_s$ in analysis B, the world average value of $\gamma$ from tree-level decays, $\gamma = (70.1 \pm 7.1)^\circ$~\cite{UTFIT}, is used as an input, and $-2\beta_s$ is left as a free parameter.

The inputs to the analyses are the measured values of $C_{\pi^+\pi^-}$, $S_{\pi^+\pi^-}$, $C_{K^+K^-}$, $S_{K^+K^-}$, $\mathcal{B}_{\pi^+\pi^-}$ and $\mathcal{B}_{K^+K^-}$. 
The corresponding constraints are given in Eqs.~\ref{eq:cpipi},~\ref{eq:spipi},~\ref{eq:ckk},~\ref{eq:skk},~\ref{eq:bpipi} and~\ref{eq:bkk}.
In addition, the value of $A^{\Delta\Gamma}_{K^+K^-}$ is fixed to be negative.
A summary of the experimental inputs is given in Table~\ref{tab:FLinputs}.

\begin{table}[t]
\begin{center}
\caption{\small Experimental inputs used for the determination of $\gamma$ and $-2\beta_s$ from $B^0 \to \pi^+\pi^-$ and $B^0_s \to K^+K^-$ decays using U-spin symmetry. The parameter $\rho(X,\,Y)$ is the statistical correlation between $X$ and $Y$. For $C_{\pi^+\pi^-}$ and $S_{\pi^+\pi^-}$ we perform our own weighted average of BaBar, Belle and LHCb results, accounting for correlations.}\label{tab:FLinputs}
%\resizebox{1\textwidth}{!}{
\begin{tabular}{c|c|l}
Quantity & Value & Source \\
\hline
$C_{\pi^+\pi^-}$ &$-0.30 \pm 0.05$ & This Letter \\
$S_{\pi^+\pi^-}$ & $-0.66 \pm 0.06$ & This Letter \\
$\mathrm{\rho}(C_{\pi^+\pi^-},\,S_{\pi^+\pi^-})$ & $-0.007$ & This Letter \\
$C_{K^+K^-}$ &$\phantom{-}0.14 \pm 0.11$ & LHCb~\cite{Aaij:2013tna} \\
$S_{K^+K^-}$ & $\phantom{-}0.30 \pm 0.13$ & LHCb~\cite{Aaij:2013tna} \\
$\mathrm{\rho}(C_{K^+K^-},\,S_{K^+K^-})$ & $\phantom{-}0.02$ & LHCb~\cite{Aaij:2013tna} \\
$\mathcal{B}_{\pi^+\pi^-}\times 10^6$ & $\phantom{-}5.10 \pm 0.19$ & HFAG~$\!\!\!$\cite{bib:hfagbase} \\
$\mathcal{B}_{K^+K^-}\times 10^6$ & $\phantom{-}24.5 \pm 1.8\phantom{0}$ & HFAG~$\!\!\!$\cite{bib:hfagbase} \\
$\sin2\beta$ & $\phantom{-}0.682 \pm 0.019$ & HFAG~$\!\!\!$\cite{bib:hfagbase} \\
$\gamma$ (analysis B only) & $\hspace{2.8mm}(70.1 \pm 7.1)^\circ$ & UTfit~~$\!$\cite{UTFIT} \\
$\lambda$ & $\hspace{3.3mm}0.2253 \pm 0.0007$ & PDG\phantom{b}~$\!\!\!$\cite{Beringer:1900zz} \\
$m_{B^0}$ $[\!\mevcc]$& $ 5279.55 \pm 0.26\phantom{0}\hspace{0.5mm} $  & PDG\phantom{b}~$\!\!\!$\cite{Beringer:1900zz} \\
$m_{B^0_s}$ $[\!\mevcc]$& $ 5366.7\pm 0.4\hspace{2.5mm} $  & PDG\phantom{b}~$\!\!\!$\cite{Beringer:1900zz} \\
$m_{\pi^+}$ $[\!\mevcc]$& $ 139.57018 \pm  0.00035 $  & PDG\phantom{b}~$\!\!\!$\cite{Beringer:1900zz} \\
$m_{K^+}$ $[\!\mevcc]$& $ 493.677 \pm  0.013 $  & PDG\phantom{b}~$\!\!\!$\cite{Beringer:1900zz} \\
$\tau_{B^0}$ $[\!\ps]$ & $ \phantom{00}1.519 \pm 0.007 $  & HFAG~$\!\!\!$\cite{bib:hfagbase} \\
$\tau_{B^0_s}$ $[\!\ps]$ & $ \phantom{00}1.516 \pm 0.011 $  & HFAG~$\!\!\!$\cite{bib:hfagbase} \\
$\Delta\Gamma_s / \Gamma_s $ & $\phantom{00}0.160 \pm 0.020 $  & LHCb~\cite{Aaij:2013oba} \\
$\tau(B^0_s \to K^+K^-)$ $[\!\ps]$ & $\phantom{00}1.452 \pm 0.042 $  & LHCb~\cite{Aaij:2012kn,Aaij:2012ns,bib:hfagbase}
\end{tabular}
\vspace{-0.6 cm}
%}
\end{center}
\end{table}

\begin{table}[t]
\begin{center}
\caption{\small Ranges of flat priors used for the determination of $\gamma$ and $-2\beta_s$ from $B^0 \to \pi^+\pi^-$ and $B^0_s \to K^+K^-$ decays using U-spin symmetry.}\label{tab:Fpriors}
%\resizebox{1\textwidth}{!}{
\begin{tabular}{c|c}
Quantity & Prior range \\
\hline
$d$ & $[0,\,20]$ \\
$\vartheta$ & $[-180^\circ,\,180^\circ]$ \\
$r_D$ & $[0,\,\kappa]$ \\
$\vartheta_{r_D}$ & $[-180^\circ,\,180^\circ]$ \\
$r_G$ & $[0,\,\kappa]$ \\
$\vartheta_{r_G}$ & $[-180^\circ,\,180^\circ]$ \\
$\gamma$ (analysis A only) & $[-180^\circ,\,180^\circ]$ \\
$-2\beta_s$ [rad] (analysis B only) & $[-\pi,\,\pi]$ 
\end{tabular}
%}
\end{center}
\end{table}

In both analyses, flat prior probability distributions, hereinafter referred to as priors, on $d$, $\vartheta$, $r_D$, $\vartheta_{r_D}$, $r_G$, $\vartheta_{r_G}$ and, where appropriate, on $\gamma$ and $-2\beta_s$ are used. In particular, we allow the U-spin breaking phases $\vartheta_{r_D}$ and $\vartheta_{r_G}$ to be completely undetermined, using flat priors between $-180^\circ$ and $180^\circ$. Concerning the parameters $r_D$ and $r_G$, we adopt uniform priors between $0$ and $\kappa$, where $\kappa$ represents the maximum magnitude of non-factorizable U-spin breaking allowed. The ranges of the flat priors are summarized in Table~\ref{tab:Fpriors}.
We study the sensitivity on $\gamma$ and $-2\beta_s$ as a function of $\kappa$, ranging from $0$ to $1$, meaning from $0\%$ up to $100\%$ non-factorizable U-spin breaking.
For all experimental inputs we use Gaussian PDFs. The values of $|D^\prime|$, $d^\prime$ and $\vartheta^\prime$ are determined using Eqs.~\ref{eq:break1} and~\ref{eq:break2}. 

\begin{figure}[t]
  \begin{center}
    \includegraphics[width=0.48\textwidth]{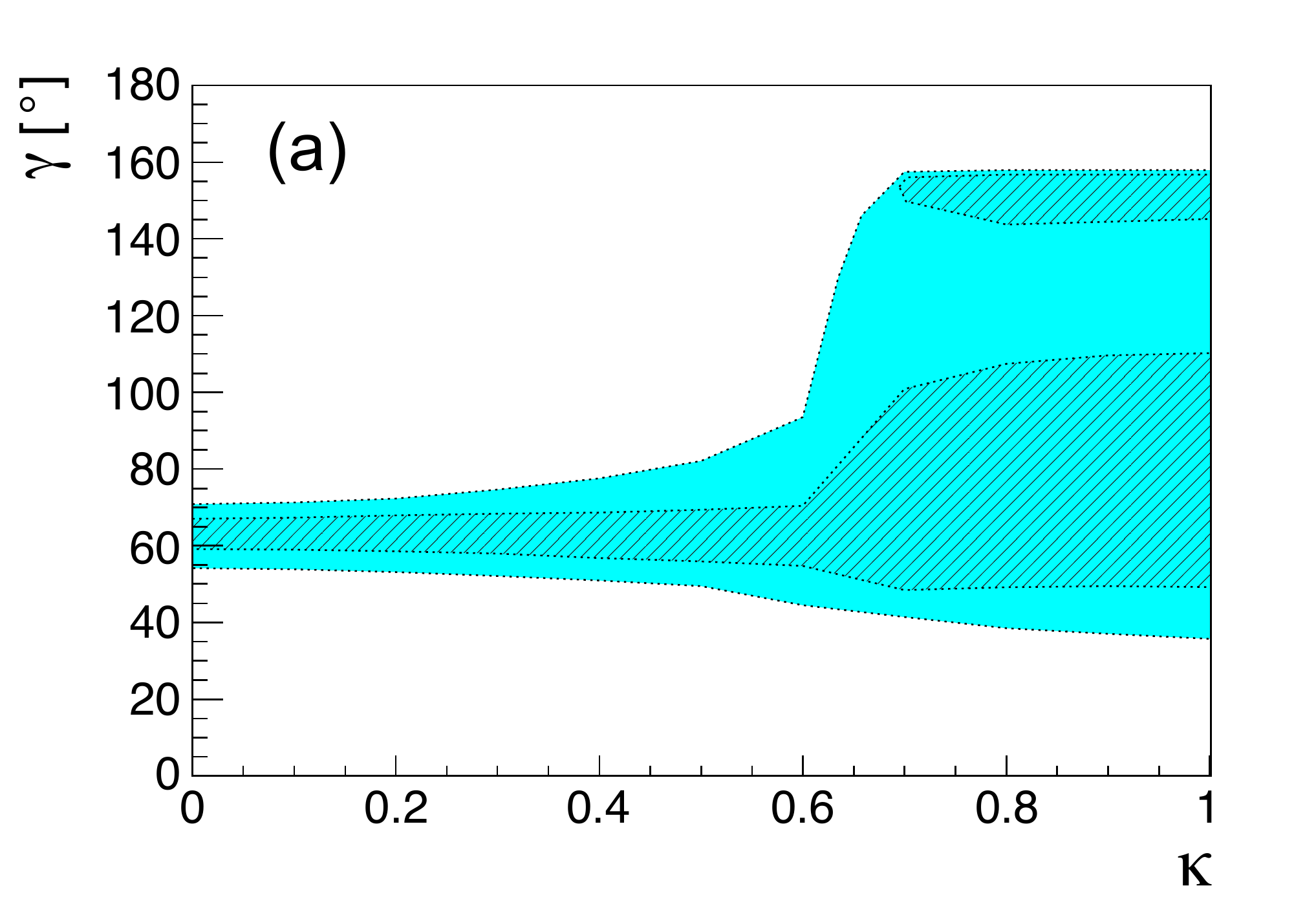}
    \includegraphics[width=0.48\textwidth]{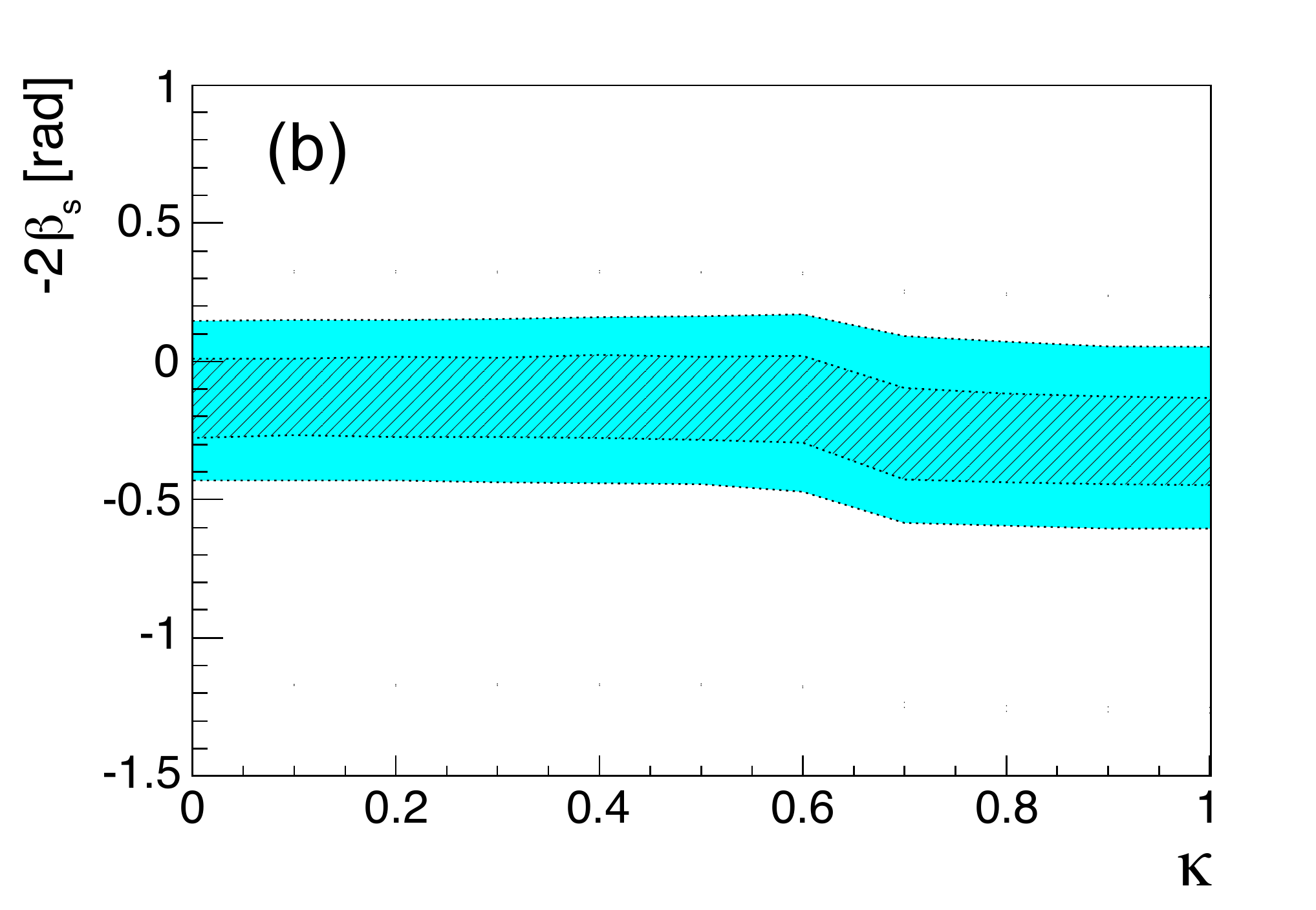}
\end{center}
  \caption{\small Dependences of the $68\%$ (hatched areas) and $95\%$ (filled areas) probability intervals on the allowed amount of non-factorizable U-spin breaking, for (a)~$\gamma$ from analysis A and (b)~$-2\beta_s$ from analysis B.}
  \label{fig:FLgphisuspinbreak}
\end{figure}

The dependences on $\kappa$ of the $68\%$ and $95\%$ posterior probability intervals for $\gamma$ and $-2\beta_s$ are shown in Fig.~\ref{fig:FLgphisuspinbreak}. When the allowed amount of U-spin breaking becomes large enough, the PDF for $\gamma$ is poorly constrained. In particular, it can be noted that for values of $\kappa$ exceeding 0.6 the sensitivity on $\gamma$ reduces significantly as a function of increasing $\kappa$. This fast transition is related to the non-linearity of the constraint equations. For $-2\beta_s$ the dependence of the sensitivity on $\kappa$ is mild, but for values of $\kappa$ exceeding 0.6 a slight shift of the distribution towards more negative values is observed.

\begin{figure}[t]
  \begin{center}
    \includegraphics[width=0.48\textwidth]{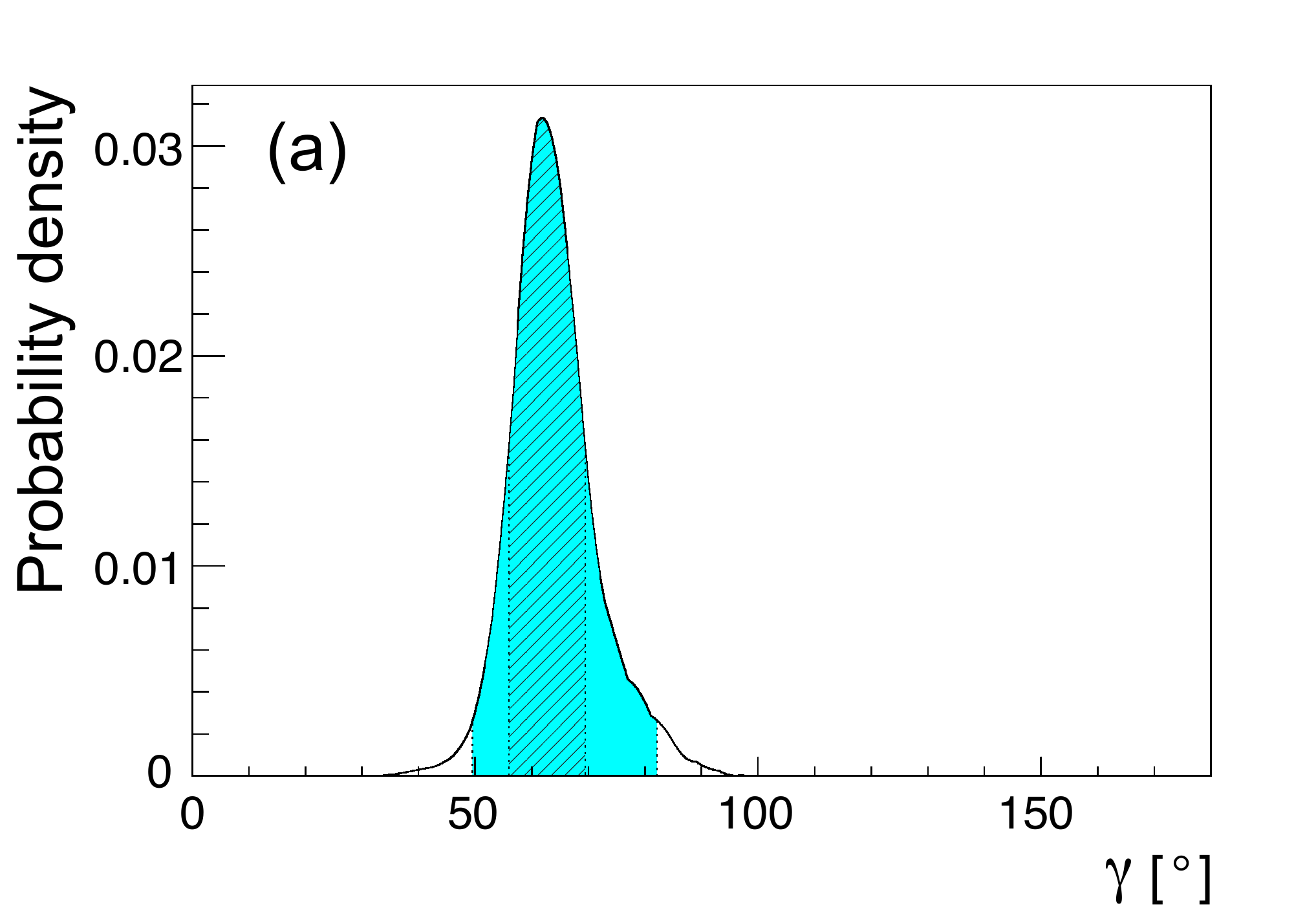}
   \includegraphics[width=0.48\textwidth]{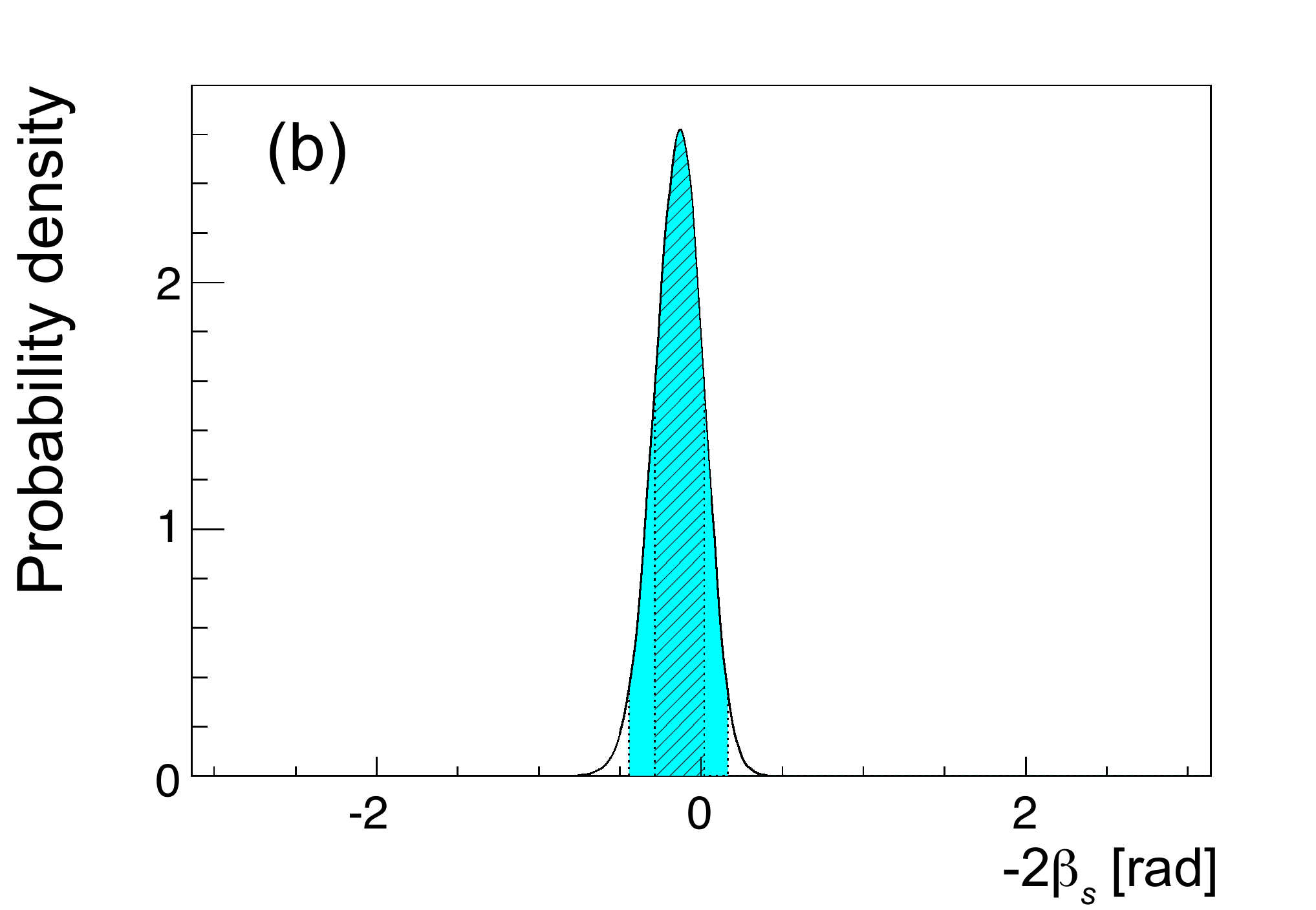}
\end{center}
\caption{\small Distributions of (a)~$\gamma$ from analysis A and (b)~$-2\beta_s$ from analysis B, corresponding to $\kappa=0.5$. The hatched areas correspond to 68\% probability intervals, whereas the filled areas correspond to 95\% probability intervals.}
\label{fig:FLgammauspinbreak100final}
\end{figure}

In Fig.~\ref{fig:FLgammauspinbreak100final} we show the PDFs for $\gamma$ obtained from analysis A and for $-2\beta_s$ obtained from analysis B, corresponding to $\kappa=0.5$. The numerical results from both analyses are reported in Table~\ref{tab:FLgammauspinbreak100resultsfinal}.
The 68\% probability interval for $\gamma$ is $[56^\circ,\,70^\circ]$, and that for $-2\beta_s$ is $[-0.28,\,0.02]$\rad.

\begin{table}[t]
\begin{center}
\caption{\small Results obtained from analyses A and B with $\kappa=0.5$. The results are given modulo $180^\circ$ for $\vartheta$, $\vartheta^\prime$ and $\gamma$.}\label{tab:FLgammauspinbreak100resultsfinal}
%\resizebox{1\textwidth}{!}{
\begin{tabular}{c|cc|cc}
             & \multicolumn{2}{c|}{Analysis A} & \multicolumn{2}{c}{Analysis B} \\
Quantity & 68\% prob. & 95\% prob.  & 68\% prob. & 95\% prob. \\
\hline
$d$ & $[0.32,\,0.53]$ & $[0.25,\,0.78]$  & $\phantom{-}[0.36,\,0.58]$ & $\phantom{-}[0.29,\,0.75]$\\
$\vartheta$ & $[136^\circ,\,157^\circ]$  & $[119^\circ,\,165^\circ]$ & $\phantom{-}[141^\circ,\,157^\circ]$  & $\phantom{-}[129^\circ,\,163^\circ]$\\
$d^\prime$ & $[0.33,\,0.50]$ & $[0.28,\,0.65]$ & $\phantom{-}[0.34,\,0.52]$ & $\phantom{-}[0.28,\,0.69]$  \\
$\vartheta^\prime$ & $[132^\circ,\,160^\circ]$  & $[114^\circ,\,176^\circ]$ & $\phantom{-}[132^\circ,\,160^\circ]$  & $\phantom{-}[117^\circ,\,175^\circ]$ \\
$|D|$ [$\!\mev^{\frac{1}{2}}\ps^{-\frac{1}{2}}$] & $[0.102,\,0.114]$ & $[0.094,\,0.121]$ & $\phantom{-}[0.101,\,0.112]$ & $\phantom{-}[0.095,\,0.117]$ \\
$|D^\prime|$ [$\!\mev^{\frac{1}{2}}\ps^{-\frac{1}{2}}$] & $[0.130,\,0.195]$ & $[0.097,\,0.231]$ & $\phantom{-}[0.122,\,0.188]$ & $\phantom{-}[0.090,\,0.224]$ \\
$\gamma$ & $[56^\circ,\,70^\circ]$ & $[49^\circ,\,82^\circ]$ & $\phantom{-}-$ & $\phantom{-}-$ \\ 
$-2\beta_s$ [$\!\rad$] & $-$ & $-$ & $[-0.28,\,0.02]$ & $[-0.44,\, 0.17]$
\end{tabular}
%}
\end{center}
\end{table}

\section{Inclusion of physics observables from {\boldmath $B^0 \to \pi^0\pi^0$} and {\boldmath $B^+ \to \pi^+\pi^0$} decays}
\label{sec:combined}

A method to determine the angle $\alpha$ of the UT using \CP asymmetries and branching fractions of $B^0 \to \pi^+\pi^-$, $B^0 \to \pi^0\pi^0$ and $B^+ \to \pi^+\pi^0$ decays was proposed in Ref.~\cite{Gronau:1990ka}. This method relies on the isospin symmetry of strong interactions and on the assumption of negligible contributions from electroweak penguin amplitudes. Isospin breaking and electroweak penguin contributions are known to be small, and their impact on the determination of the weak phase is at the level of $1^\circ$~\cite{Buras:1998rb,Gronau:1998fn,Zupan:2007fq,Botella:2006zi}. In Ref.~\cite{Ciuchini:2012gd} it was suggested to combine the isospin-based technique of Ref.~\cite{Gronau:1990ka} with that of Ref.~\cite{Fleischer:1999pa} based on U-spin. Here we extend the study presented in Sec.~\ref{sec:FL} by including the experimental information on $B^0 \to \pi^0\pi^0$ and $B^+ \to \pi^+\pi^0$ decays, \emph{i.e.} using also the observables $C_{\pi^0\pi^0}$, $\mathcal{B}_{\pi^0\pi^0}$ and $\mathcal{B}_{\pi^+\pi^0}$. 
The corresponding constraints are given in Eqs.~\ref{eq:cpi0pi0},~\ref{eq:bpi0pi0} and~\ref{eq:bpipi0}.

In complete analogy with the study presented in Sec.~\ref{sec:FL}, we perform two distinct analyses, to determine either $\gamma$ or $-2\beta_s$. They are referred to as analyses C and D, respectively. In analysis C, the value of $-2\beta_s$ is constrained as a function of $\beta$ and $\gamma$, and $\gamma$ is determined, whereas in analysis D, the world average value of $\gamma$ from tree-level decays is used as an input and $-2\beta_s$ is determined.  A summary of the experimental inputs is given in Table~\ref{tab:Cinputs}. 

\begin{table}[tb]
\begin{center}
\caption{\small Experimental inputs used for the determination of $\gamma$ and $-2\beta_s$ from $B^0 \to \pi^+\pi^-$, $B^0 \to \pi^0\pi^0$, $B^+ \to \pi^+\pi^0$ and $B^0_s \to K^+K^-$  decays, using isospin and U-spin symmetries. The parameter $\rho(X,\,Y)$ is the statistical correlation between $X$ and $Y$. For $C_{\pi^+\pi^-}$ and $S_{\pi^+\pi^-}$ we perform our own weighted average of BaBar, Belle and LHCb results, accounting for correlations.}\label{tab:Cinputs}
%\resizebox{1\textwidth}{!}{
\begin{tabular}{c|c|l}
Quantity & Value & Source \\
\hline
$C_{\pi^+\pi^-}$ &$-0.30 \pm 0.05$ & This Letter \\
$S_{\pi^+\pi^-}$ & $-0.66 \pm 0.06$ & This Letter \\
$\mathrm{\rho}(C_{\pi^+\pi^-},\,S_{\pi^+\pi^-})$ & $-0.007$ & This Letter \\
$C_{\pi^0\pi^0}$ & $-0.43 \pm 0.24$ & HFAG~$\!\!\!$\cite{bib:hfagbase} \\
$C_{K^+K^-}$ &$\phantom{-}0.14 \pm 0.11$ & LHCb~\cite{Aaij:2013tna} \\
$S_{K^+K^-}$ & $\phantom{-}0.30 \pm 0.13$ & LHCb~\cite{Aaij:2013tna} \\
$\mathrm{\rho}(C_{K^+K^-},\,S_{K^+K^-})$ & $\phantom{-}0.02$ & LHCb~\cite{Aaij:2013tna} \\
$\mathcal{B}_{\pi^+\pi^-}\times 10^6$ & $\phantom{-}5.10 \pm 0.19$ & HFAG~$\!\!\!$\cite{bib:hfagbase} \\
$\mathcal{B}_{\pi^+\pi^0}\times 10^6$ & $\phantom{-}5.48 \pm 0.35$ & HFAG~$\!\!\!$\cite{bib:hfagbase} \\
$\mathcal{B}_{\pi^0\pi^0}\times 10^6$ & $\phantom{-}1.91 \pm 0.23$  & HFAG~$\!\!\!$\cite{bib:hfagbase} \\
$\mathcal{B}_{K^+K^-}\times 10^6$ & $\phantom{-}24.5 \pm 1.8\phantom{0}$ & HFAG~$\!\!\!$\cite{bib:hfagbase} \\
$\sin2\beta$ & $\phantom{00}\! 0.682 \pm 0.019$ & HFAG~$\!\!\!$\cite{bib:hfagbase} \\
$\gamma$ (analysis D only) & $\hspace{2.8mm} (70.1 \pm 7.1)^\circ$ & UTfit~~$\!$\cite{UTFIT} \\
$\lambda$ & $\hspace{3.3mm} 0.2253 \pm 0.0007$ & PDG\phantom{b}~$\!\!\!$\cite{Beringer:1900zz} \\
$m_{B^0}$ $[\!\mevcc]$& $ 5279.55 \pm 0.26\phantom{0}\hspace{0.5mm} $  & PDG\phantom{b}~$\!\!\!$\cite{Beringer:1900zz} \\
$m_{B^+}$ $[\!\mevcc]$& $ 5279.25 \pm 0.26\phantom{0}\hspace{0.5mm} $  & PDG\phantom{b}~$\!\!\!$\cite{Beringer:1900zz} \\
$m_{B^0_s}$ $[\!\mevcc]$& $ 5366.7\pm 0.4\hspace{2.5mm} $  & PDG\phantom{b}~$\!\!\!$\cite{Beringer:1900zz} \\
$m_{\pi^+}$ $[\!\mevcc]$& $ 139.57018 \pm  0.00035 $  & PDG\phantom{b}~$\!\!\!$\cite{Beringer:1900zz} \\
$m_{\pi^0}$ $[\!\mevcc]$& $ 134.9766 \pm 0.0006 $  & PDG\phantom{b}~$\!\!\!$\cite{Beringer:1900zz} \\
$m_{K^+}$ $[\!\mevcc]$& $ 493.677 \pm  0.013 $  & PDG\phantom{b}~$\!\!\!$\cite{Beringer:1900zz} \\
$\tau_{B^0}$ $[\!\ps]$ & $\phantom{00} 1.519 \pm 0.007 $  & HFAG~$\!\!\!$\cite{bib:hfagbase} \\
$\tau_{B^+}$ $[\!\ps]$ & $\phantom{00} 1.641 \pm 0.008 $  & HFAG~$\!\!\!$\cite{bib:hfagbase} \\ 
$\tau_{B^0_s}$ $[\!\ps]$ & $\phantom{00} 1.516 \pm 0.011 $  & HFAG~$\!\!\!$\cite{bib:hfagbase} \\
$\Delta\Gamma_s / \Gamma_s $ & $\phantom{00} 0.160 \pm 0.020 $  & LHCb~\cite{Aaij:2013oba} \\
$\tau(B^0_s \to K^+K^-)$ $[\!\ps]$ & $\phantom{00} 1.452 \pm 0.042 $  & LHCb~\cite{Aaij:2012kn,Aaij:2012ns,bib:hfagbase} 
\vspace{-0.5cm}
\end{tabular}
%}
\end{center}
\end{table}

In both analyses, flat priors on $d$, $\vartheta$, $q$, $\vartheta_q$, $r_D$, $\vartheta_{r_D}$, $r_G$, $\vartheta_{r_G}$ and, where appropriate, on $\gamma$ and $-2\beta_s$ are used. The ranges of the flat priors are summarized in Table~\ref{tab:Cpriors}. For all experimental inputs we use Gaussian PDFs. The values of $|D^\prime|$, $d^\prime$ and $\vartheta^\prime$ are again determined using Eqs.~\ref{eq:break1} and~\ref{eq:break2}.

\begin{table}[tb]
\begin{center}
\caption{\small Ranges of flat priors used for the determination of $\gamma$ and $-2\beta_s$ from $B^0 \to \pi^+\pi^-$, $B^0 \to \pi^0\pi^0$, $B^+ \to \pi^+\pi^0$ and $B^0_s \to K^+K^-$  decays, using isospin and U-spin symmetries.}\label{tab:Cpriors}
%\resizebox{1\textwidth}{!}{
\begin{tabular}{c|c}
Quantity & Prior range \\
\hline
$d$ & $[0,\,20]$ \\
$\vartheta$ & $[-180^\circ,\,180^\circ]$ \\
$q$  & $[0,\,20]$ \\
$\vartheta_q$ & $[-180^\circ,\,180^\circ]$ \\
$r_D$ & $[0,\,\kappa]$ \\
$\vartheta_{r_D}$ & $[-180^\circ,\,180^\circ]$ \\
$r_G$ & $[0,\,\kappa]$ \\
$\vartheta_{r_G}$ & $[-180^\circ,\,180^\circ]$ \\
$\gamma$ (analysis C only) & $[-180^\circ,\,180^\circ]$ \\
$-2\beta_s$ [rad] (analysis D only) & $[-\pi,\,\pi]$ 
\end{tabular}
%}
\end{center}
\end{table}

\begin{figure}[tb]
  \begin{center}
    \includegraphics[width=0.48\textwidth]{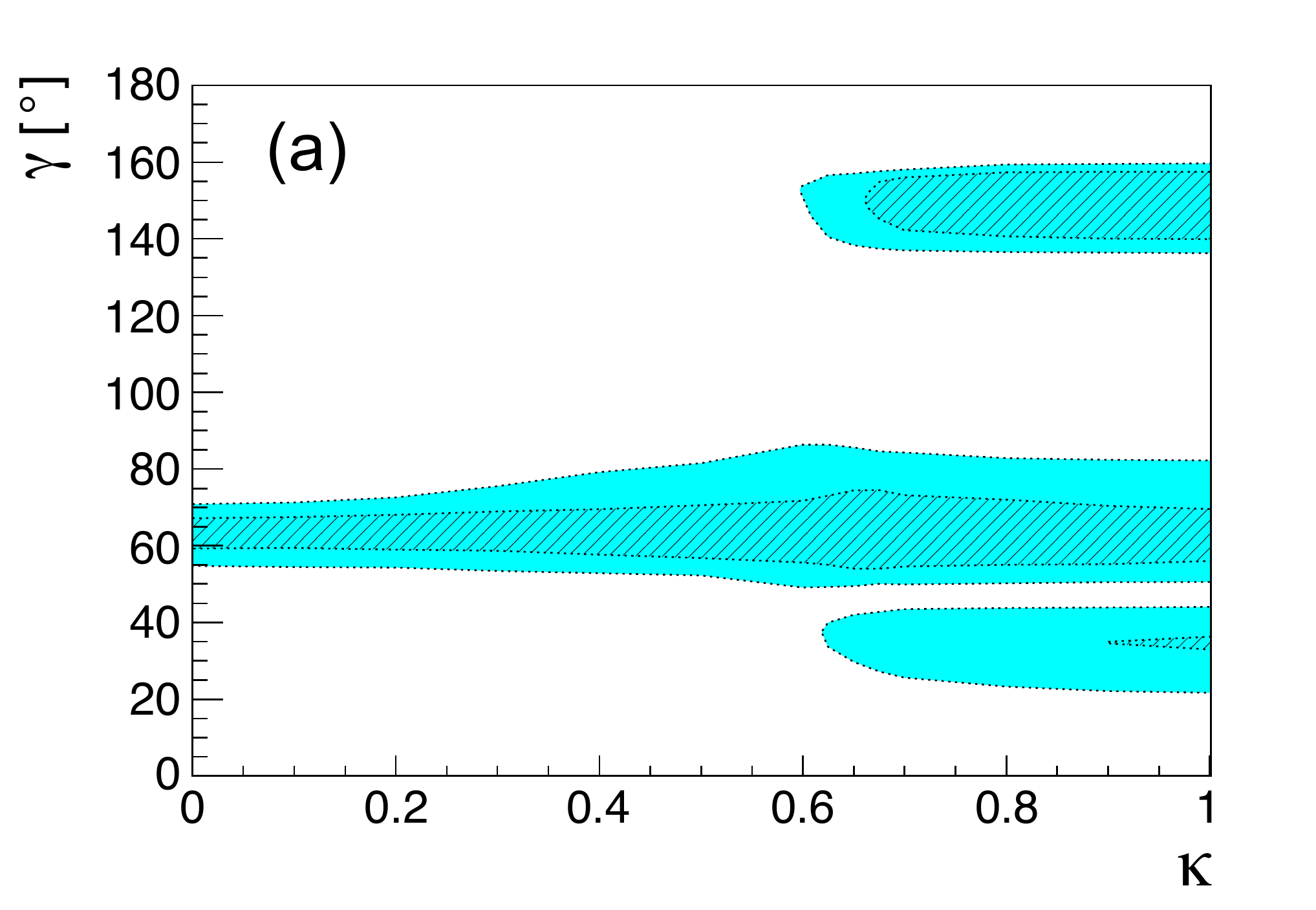}
    \includegraphics[width=0.48\textwidth]{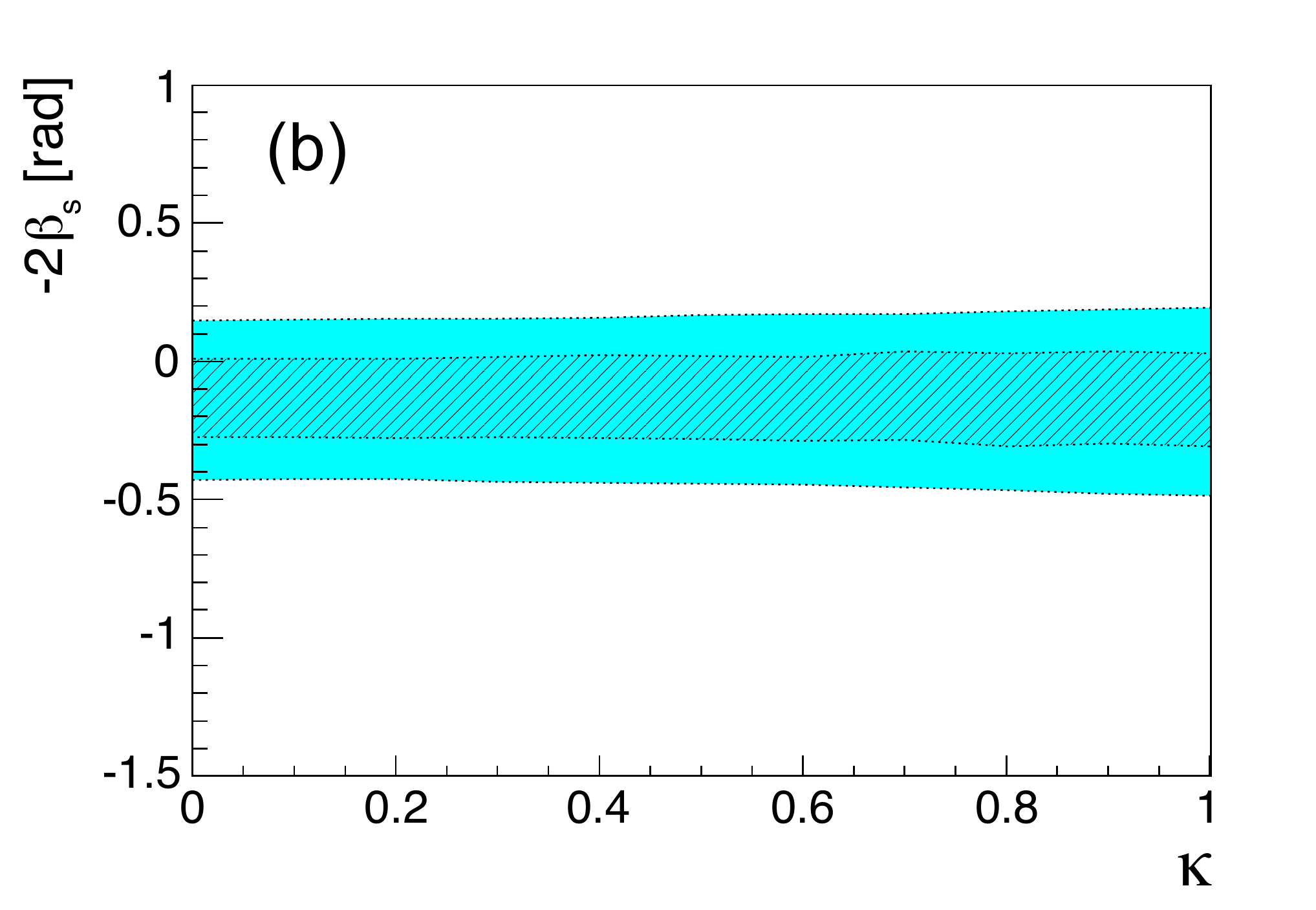}
\end{center}
  \caption{\small Dependences of the $68\%$ (hatched areas) and $95\%$ (filled areas) probability intervals on the allowed amount of non-factorizable U-spin breaking, for (a)~$\gamma$ from analysis C and (b)~$-2\beta_s$ from analysis D.}
  \label{fig:Cgphisuspinbreak}
\end{figure}

\begin{figure}[tb]
  \begin{center}
   \includegraphics[width=0.48\textwidth]{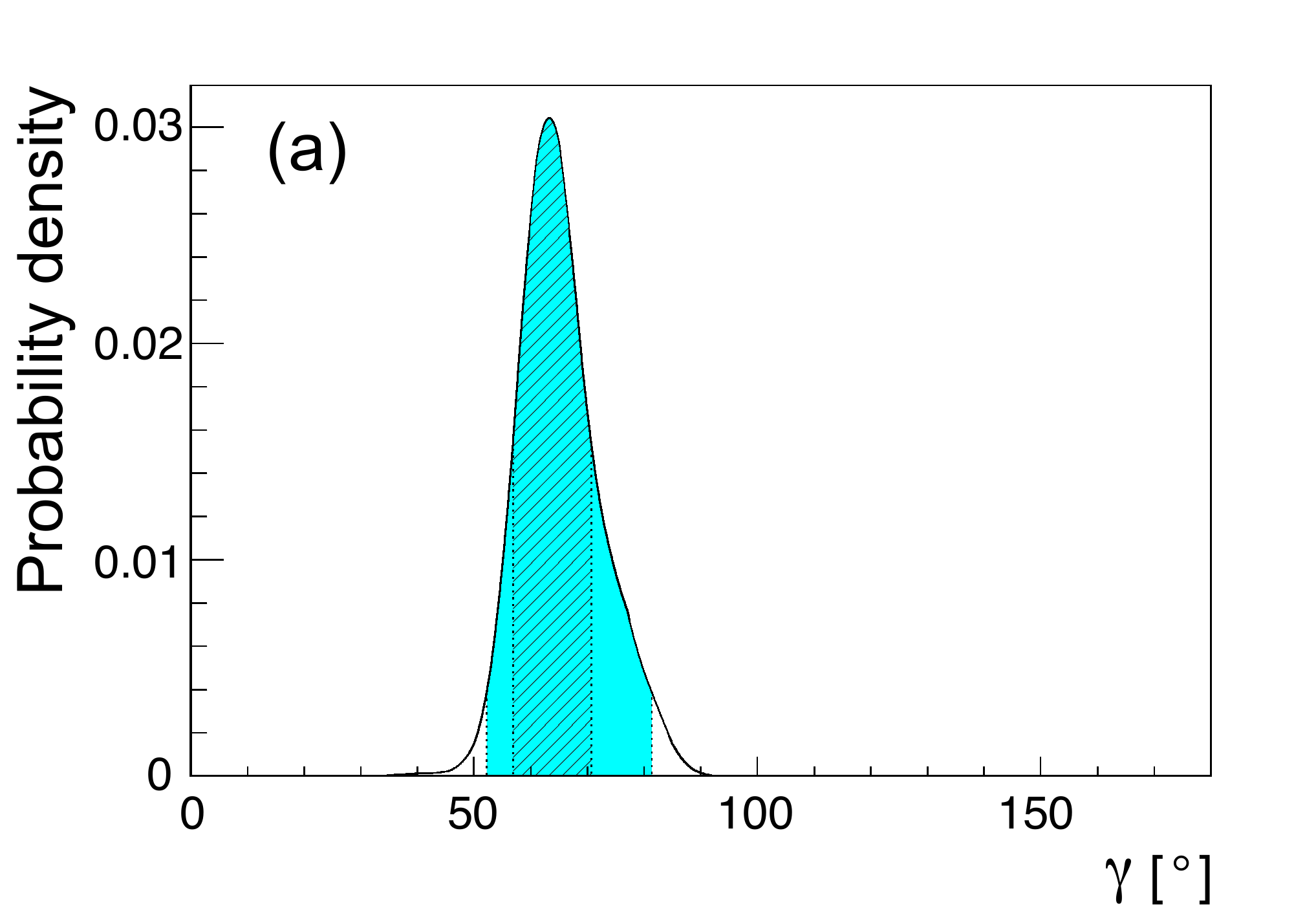}
   \includegraphics[width=0.48\textwidth]{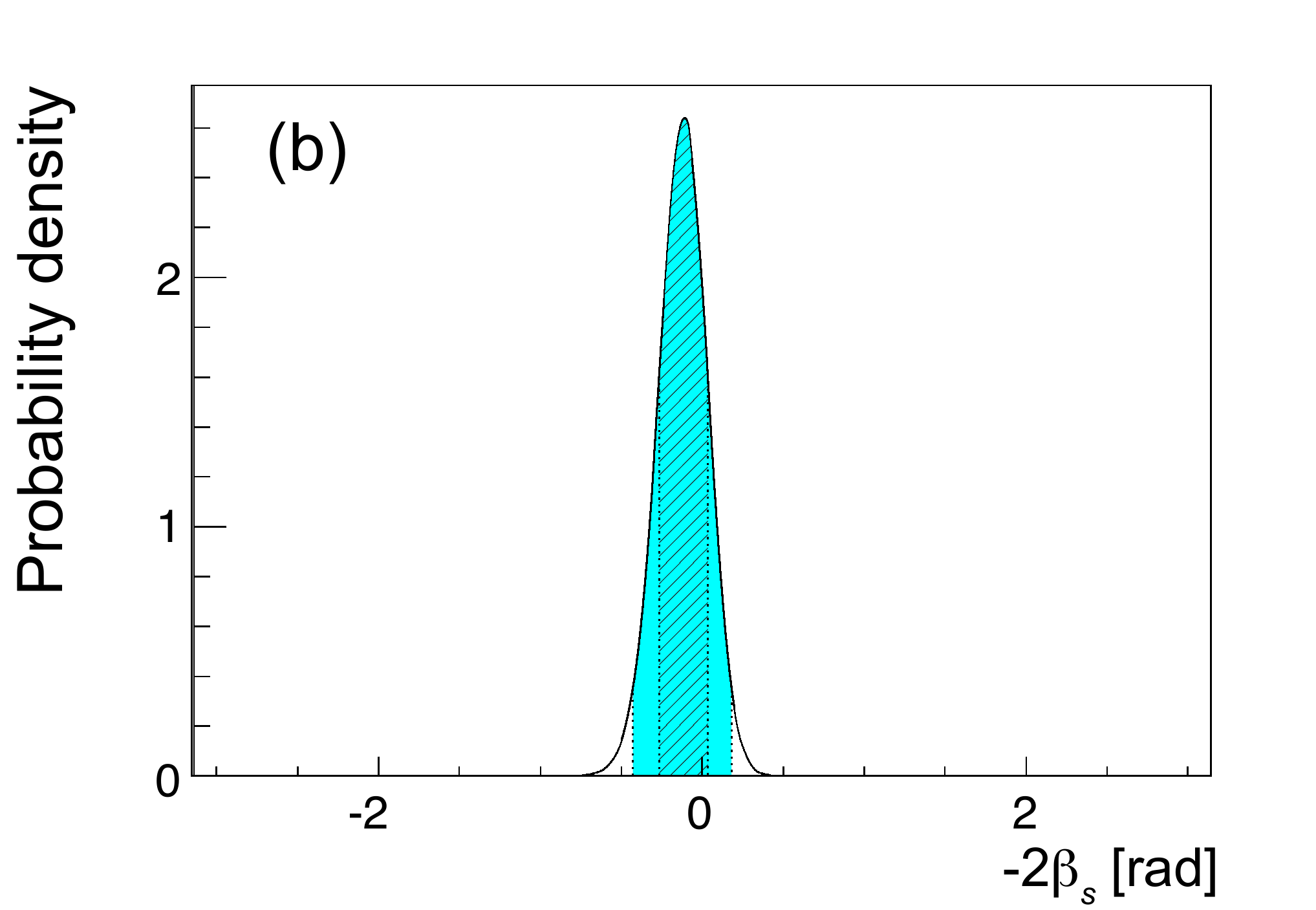}
\end{center}
\caption{\small Distributions of (a)~$\gamma$ from analysis C and (b)~$-2\beta_s$ from analysis D, corresponding to $\kappa=0.5$. The hatched areas correspond to 68\% probability intervals, whereas the filled areas correspond to 95\% probability intervals.}
\label{fig:Cgammauspinbreak100final}
\end{figure}

The dependences on $\kappa$ of the $68\%$ and $95\%$ probability intervals for $\gamma$ and $-2\beta_s$ are shown in Fig.~\ref{fig:Cgphisuspinbreak}. Again, when the amount of U-spin breaking exceeds 60\%, additional maxima appear in the posterior PDF for $\gamma$. By contrast, for $-2\beta_s$, the dependence of the sensitivity on $\kappa$ is very weak.
In Fig.~\ref{fig:Cgammauspinbreak100final} we show the PDFs for $\gamma$ obtained from analysis C and for $-2\beta_s$ obtained from analysis D, corresponding to $\kappa=0.5$. The numerical results from both analyses are reported in Table~\ref{tab:Cgammauspinbreak100resultsfinal}.
The 68\% probability interval for $\gamma$ is $[57^\circ,\,71^\circ]$, and that for $-2\beta_s$ is $[-0.28,\,0.02]$\rad.

%\clearpage

%\clearpage

\begin{table}[tb]
\begin{center}
\caption{\small Results obtained from analyses C and D with $\kappa=0.5$. The results are given modulo $180^\circ$ for $\vartheta$, $\vartheta^\prime$ and $\gamma$.}\label{tab:Cgammauspinbreak100resultsfinal}
%\resizebox{1\textwidth}{!}{
\begin{tabular}{c|cc|cc}
             & \multicolumn{2}{c|}{Analysis C} & \multicolumn{2}{c}{Analysis D} \\
Quantity & 68\% prob. & 95\% prob.  & 68\% prob. & 95\% prob. \\
\hline
$d$ & $[0.33,\,0.57]$ & $[0.28,\,0.79]$  & $\phantom{-}[0.37,\,0.59]$ & $\phantom{-}[0.31,\,0.77]$\\
$\vartheta$ & $[139^\circ,\,157^\circ]$  & $[125^\circ,\,164^\circ]$ & $\phantom{-}[142^\circ,\,157^\circ]$  & $\phantom{-}[132^\circ,\,163^\circ]$\\
$d^\prime$ & $[0.34,\,0.50]$ & $[0.28,\,0.65]$ & $\phantom{-}[0.34,\,0.52]$ & $\phantom{-}[0.29,\,0.70]$  \\
$\vartheta^\prime$ & $[132^\circ,\,160^\circ]$  & $[119^\circ,\,176^\circ]$  & $\phantom{-}[133^\circ,\,160^\circ]$  & $\phantom{-}[119^\circ,\,176^\circ]$ \\
$q$  & $[1.04,\,1.21]$ & $[0.94,\,1.30]$ & $\phantom{-}[1.04,\,1.21]$ & $\phantom{-}[0.95,\,1.30]$ \\
$\vartheta_q$ & $[-82^\circ,\,-58^\circ]$ & $[-88^\circ,\,-35^\circ]$ & $\phantom{-}[-78^\circ,\,-57^\circ]$ & $[-85^\circ,\-38^\circ]$ \\
$|D|$ [$\!\mev^{\frac{1}{2}}\ps^{-\frac{1}{2}}$] & $[0.101,\,0.113]$ & $[0.094,\,0.118]$ & $\phantom{-}[0.100,\,0.111]$ & $\phantom{-}[0.094,\,0.116]$ \\
$|D^\prime|$ [$\!\mev^{\frac{1}{2}}\ps^{-\frac{1}{2}}$] & $[0.129,\,0.193]$ & $[0.097,\,0.228]$ & $\phantom{-}[0.122,\,0.187]$ & $\phantom{-}[0.089,\,0.221]$ \\
$\gamma$ & $[57^\circ,\,71^\circ]$ & $[52^\circ,\,82^\circ]$ & $\phantom{-}-$ & $\phantom{-}-$ \\
$-2\beta_s$ [$\!\rad$] & $-$ & $-$ & $[-0.28,\,0.02]$ & $[-0.44,\, 0.17]$ 
\end{tabular}
%}
\end{center}
\end{table}

It is worth emphasising that, although this study is similar to that presented in Ref.~\cite{Ciuchini:2012gd}, there are two relevant differences, in addition to the use of updated experimental inputs. First, the upper limits of the priors on $d$ and $q$ are chosen to be much larger, to include all nonzero likelihood regions and to remove any sizable dependence of the results on the choice of the priors. In particular, this leads to a bigger impact of U-spin breaking effects at very large $\kappa$ values. Second, the adopted parameterization of non-factorizable U-spin breaking is slightly different, in order to propagate equally the effects of the breaking on every topology contributing to the total decay amplitudes.

%\clearpage

\section{Results and conclusions}
\label{sec:conclusions}

Using the latest LHCb measurements of time-dependent $C\!P$ violation in the $B^0_s \to K^+K^-$ decay, and following the approaches outlined in Refs.~\cite{Fleischer:1999pa,Ciuchini:2012gd}, the angle $\gamma$ of the unitarity triangle and the $B^0_s$ mixing phase $-2\beta_s$ have been determined. The approach of Ref.~\cite{Fleischer:1999pa} relies on the use of the U-spin symmetry of strong interactions relating \BsToKK with $B^0\to \pi^+\pi^-$ decay amplitudes, whereas that of Ref.~\cite{Ciuchini:2012gd} relies on both isospin and U-spin symmetries by combining the methods proposed in Refs.~\cite{Fleischer:1999pa} and~\cite{Gronau:1990ka}, \emph{i.e.} considering also the information from $B^0\to \pi^0\pi^0$ and $B^+\to \pi^+\pi^0$ decays. To follow the latter approach, measurements solely coming from other experiments have been included in the analysis.

\begin{figure}[h!]
  \begin{center}
    \includegraphics[width=1\textwidth]{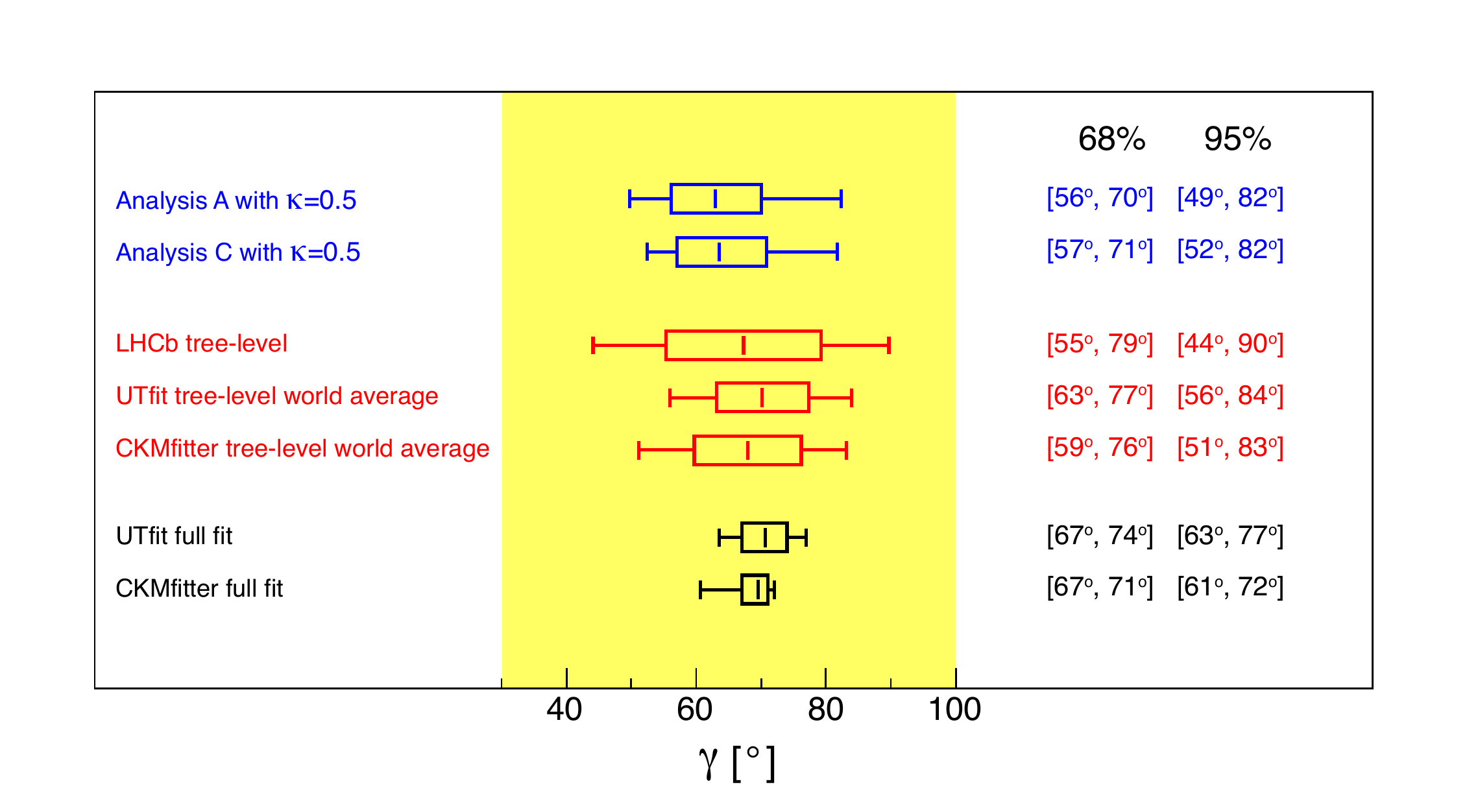}
    \includegraphics[width=1\textwidth]{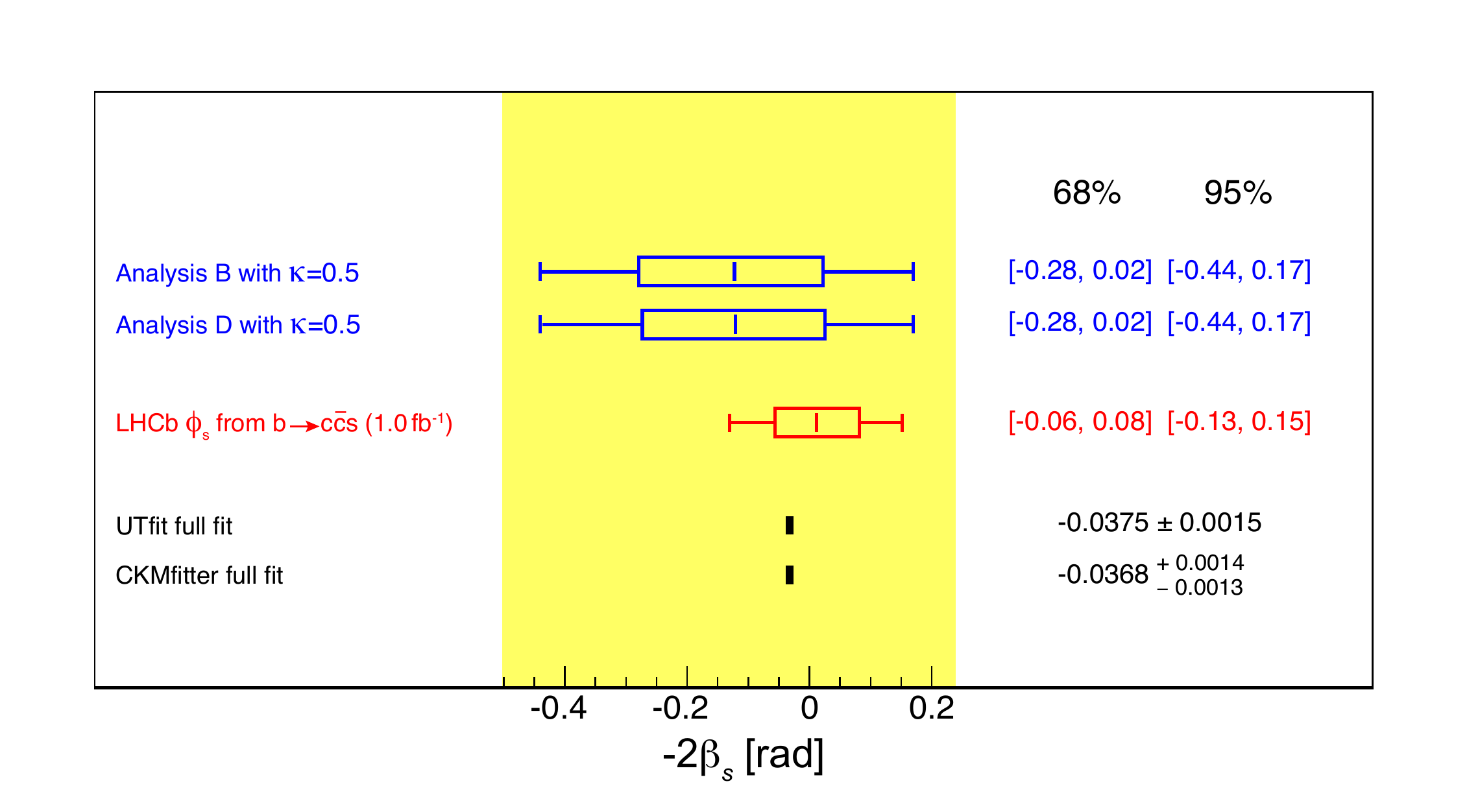}
\end{center}
  \caption{\small Results for (top) $\gamma$ and (bottom) $-2\beta_s$ with 50\% ($\kappa=0.5$) non-factorizable U-spin breaking. As a comparison, other reference values are also reported. The most likely values are indicated by the vertical lines insides the boxes. The boxes and the error bars delimit the 68\% and 95\% probability intervals, respectively.}
  \label{fig:summaryofresults}
\end{figure}

We have studied the impact of large non-factorizable U-spin breaking corrections on the determination of $\gamma$ and $-2\beta_s$. The relevant results in terms of 68\% and 95\% probability intervals, which include uncertainties due to non-factorizable U-spin breaking effects up to 50\%, are summarized in Fig.~\ref{fig:summaryofresults}. Typical U-spin breaking effects, including factorizable contributions, are expected to be much smaller, around the 30\% level~\cite{Gronau:2013mda,Nagashima:2007qn}.

With up to 50\% non-factorizable U-spin breaking, the approach of Ref.~\cite{Ciuchini:2012gd} gives marginal improvements in precision with respect to that of Ref.~\cite{Fleischer:1999pa}. The former approach gives considerably more robust results for larger U-spin breaking values. Following the approach of Ref.~\cite{Ciuchini:2012gd} and taking the most probable value as central value, at 68\% probability we obtain
\begin{equation}
\gamma =  \left( 63.5 ^{\,+\, 7.2}_{\,-\,6.7} \right)^\circ,\nonumber
\end{equation}
and, in an alternative analysis, 
\begin{equation}
-2\beta_s = -0.12 ^{\,+\,0.14}_{\,-\,0.16}\,\,\mathrm{rad}.\nonumber
\end{equation}
These results have been verified to be robust with respect to the choice of the priors and of the parameterization of non-factorizable U-spin breaking contributions.
The value of $\gamma$ shows no significant deviation from the averages of $\gamma$ from tree-level decays provided by the UTfit collaboration and the CKMfitter group that quote $\gamma = (70.1 \pm 7.1)^\circ$ and $\gamma = \left(68.0 ^{+8.0}_{-8.5} \right)^\circ$, respectively~\cite{UTFIT,CKMFITTER}. Analogously, the value of $-2\beta_s$ is compatible with the LHCb result from $b \to c\bar{c}s$ transitions, $\phi_s = 0.01 \pm 0.07\,(\mathrm{stat})\pm 0.01\,(\mathrm{syst})\,\mathrm{rad}$~\cite{Aaij:2013oba}, obtained using a data sample of $pp$ collisions corresponding to an integrated luminosity of $1.0\invfb$.

In summary, the value of $\gamma$ from charmless two-body decays of beauty mesons is found to be compatible and competitive with that from tree-level decays. However, since the impact of U-spin breaking corrections is significant, further improvements in the measurement of $\gamma$ are primarily limited by theoretical understanding of U-spin breaking. By contrast, the impact of U-spin breaking effects on the value of $-2\beta_s$ is small, and significant improvements are anticipated with the advent of larger samples of data. It is worth emphasising that the information on $-2\beta_s$ comes solely from the measurement of \CP violation in the $B^0_s \to K^+K^-$ decay~\cite{Aaij:2013tna}, also based on a data sample of $pp$ collisions corresponding to an integrated luminosity of $1.0\invfb$. At present, the overall uncertainty on $-2\beta_s$, which also includes theoretical uncertanties, is only two times larger than that obtained using $b \to c\bar{c}s$ transitions, as reported above.

%\clearpage

\section*{Acknowledgements}
\noindent We express our gratitude to our colleagues in the CERN
accelerator departments for the excellent performance of the LHC. We
thank the technical and administrative staff at the LHCb
institutes. We acknowledge support from CERN and from the national
agencies: CAPES, CNPq, FAPERJ and FINEP (Brazil); NSFC (China);
CNRS/IN2P3 (France); BMBF, DFG, HGF and MPG (Germany); SFI (Ireland); INFN (Italy); 
FOM and NWO (The Netherlands); MNiSW and NCN (Poland); MEN/IFA (Romania); 
MinES and FANO (Russia); MinECo (Spain); SNSF and SER (Switzerland); 
NASU (Ukraine); STFC (United Kingdom); NSF (USA).
The Tier1 computing centres are supported by IN2P3 (France), KIT and BMBF 
(Germany), INFN (Italy), NWO and SURF (The Netherlands), PIC (Spain), GridPP 
(United Kingdom).
We are indebted to the communities behind the multiple open 
source software packages on which we depend. We are also thankful for the 
computing resources and the access to software R\&D tools provided by Yandex LLC (Russia).
Individual groups or members have received support from 
EPLANET, Marie Sk\l{}odowska-Curie Actions and ERC (European Union), 
Conseil g\'{e}n\'{e}ral de Haute-Savoie, Labex ENIGMASS and OCEVU, 
R\'{e}gion Auvergne (France), RFBR (Russia), XuntaGal and GENCAT (Spain), Royal Society and Royal
Commission for the Exhibition of 1851 (United Kingdom).

%\addcontentsline{toc}{section}{References}
\bibliographystyle{LHCb}
\bibliography{main}

\newpage

% Author List ----------------------------
%%%%%%%%%%%%%%%%%%%%%%%%%%%%%%%%%%%%%%%%%%
\centerline{\large\bf LHCb collaboration}
\begin{flushleft}
\small
R.~Aaij$^{41}$, 
C.~Abell\'{a}n~Beteta$^{40}$, 
B.~Adeva$^{37}$, 
M.~Adinolfi$^{46}$, 
A.~Affolder$^{52}$, 
Z.~Ajaltouni$^{5}$, 
S.~Akar$^{6}$, 
J.~Albrecht$^{9}$, 
F.~Alessio$^{38}$, 
M.~Alexander$^{51}$
% ...

{\footnotesize \it
% ...
$ ^{5}$Clermont Universit\'{e}, Universit\'{e} Blaise Pascal, CNRS/IN2P3, LPC, Clermont-Ferrand, France\\
$ ^{6}$CPPM, Aix-Marseille Universit\'{e}, CNRS/IN2P3, Marseille, France\\
% ...
$ ^{9}$Fakult\"{a}t Physik, Technische Universit\"{a}t Dortmund, Dortmund, Germany\\
% ...
$ ^{37}$Universidad de Santiago de Compostela, Santiago de Compostela, Spain\\
$ ^{38}$European Organization for Nuclear Research (CERN), Geneva, Switzerland\\
% ...
$ ^{40}$Physik-Institut, Universit\"{a}t Z\"{u}rich, Z\"{u}rich, Switzerland\\
$ ^{41}$Nikhef National Institute for Subatomic Physics, Amsterdam, The Netherlands\\
% ...
$ ^{46}$H.H. Wills Physics Laboratory, University of Bristol, Bristol, United Kingdom\\
% ...
$ ^{51}$School of Physics and Astronomy, University of Glasgow, Glasgow, United Kingdom\\
$ ^{52}$Oliver Lodge Laboratory, University of Liverpool, Liverpool, United Kingdom\\
% ...
}
\end{flushleft}
%%%%%%%%%%%%%%%%%%%%%%%%%%%%%%%%%%%%%%%%%%

\end{document}